\documentclass[useAMS,usenatbib,usegraphicx]{mn2e}
\usepackage{url}

\font\japit = cmti10 at 10truept

\title
     [Covariance of Power]
{\vglue-3.0truecm
\centerline{\japit For submission to Monthly Notices}
\vglue 2.5truecm
\noindent
On measuring the covariance matrix of the nonlinear power spectrum from simulations
\author[A. J. S. Hamilton, C. D. Rimes, and R. Scoccimarro]
    {Andrew J. S. Hamilton$^{1,2}$$^\star$, Christopher D. Rimes$^1$$^\star$, and Rom\'an Scoccimarro$^{3}$\thanks{
	E-mail:
	Andrew.Hamilton@colorado.edu,
	rimes@colorado.edu,
	rs123@nyu.edu
        } \\
	$^1$JILA, University of Colorado, 440 UCB, Boulder, CO 80309-0440, USA\\
	$^2$Department of Astrophysical \& Planetary Sciences, University of Colorado, 391 UCB, Boulder, CO 80309-0391, USA\\
	$^3$Center for Cosmology and Particle Physics, Department of Physics, New York University, New York, NY 10003, USA}
}

\newcommand{\simpropto}{
  \!\begin{array}{c}
    {\propto} \\
    [-1.7ex] \sim
  \end{array}\!
}

\newcommand{\dd}{{\rmn d}}	
\newcommand{\e}{{\rmn e}}	
\newcommand{\im}{{\rmn i}}	

\newcommand{\ddd}{\dd^3}

\newcommand{\Mpc}{{\rmn{Mpc}}}

\newcommand{\bk}{{\bmath k}}

\newcommand{\br}{{\bmath r}}

\newcommand{\zero}{{\bmath 0}}

\newcommand{\rhobar}{\overline{\rho}}
\newcommand{\phat}{\hat{p}}
\newcommand{\Phat}{\widehat{P}}
\newcommand{\Xhat}{\widehat{X}}

\newcommand{\aj}[2]{AJ, #1, #2}

\newcommand{\apj}[2]{ApJ, #1, #2}
\newcommand{\apjs}[2]{ApJS, #1, #2}

\newcommand{\mn}[2]{MNRAS, #1, #2}
\newcommand{\pasp}[2]{PASP, #1, #2}

\newcommand{\prd}[2]{Phys.\ Rev.\ D, #1, #2}

\newcommand{\wfig}{
    \begin{figure}
    \begin{center}
    \leavevmode
    \includegraphics[width=2.75in]{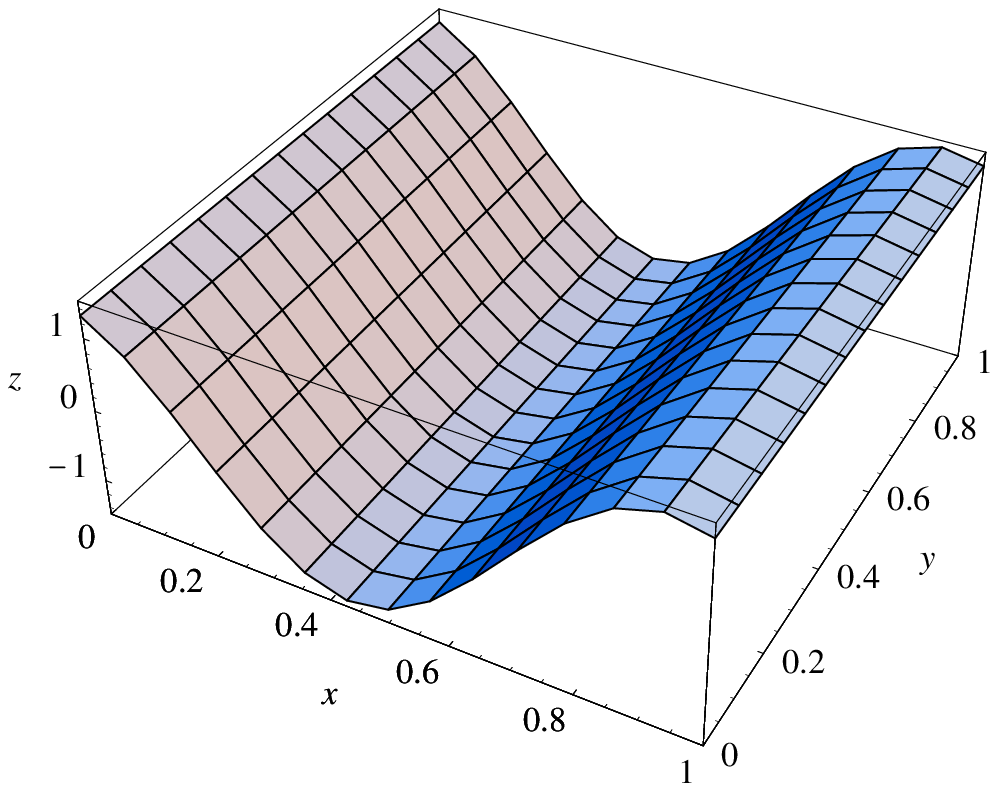}
    \includegraphics[width=2.75in]{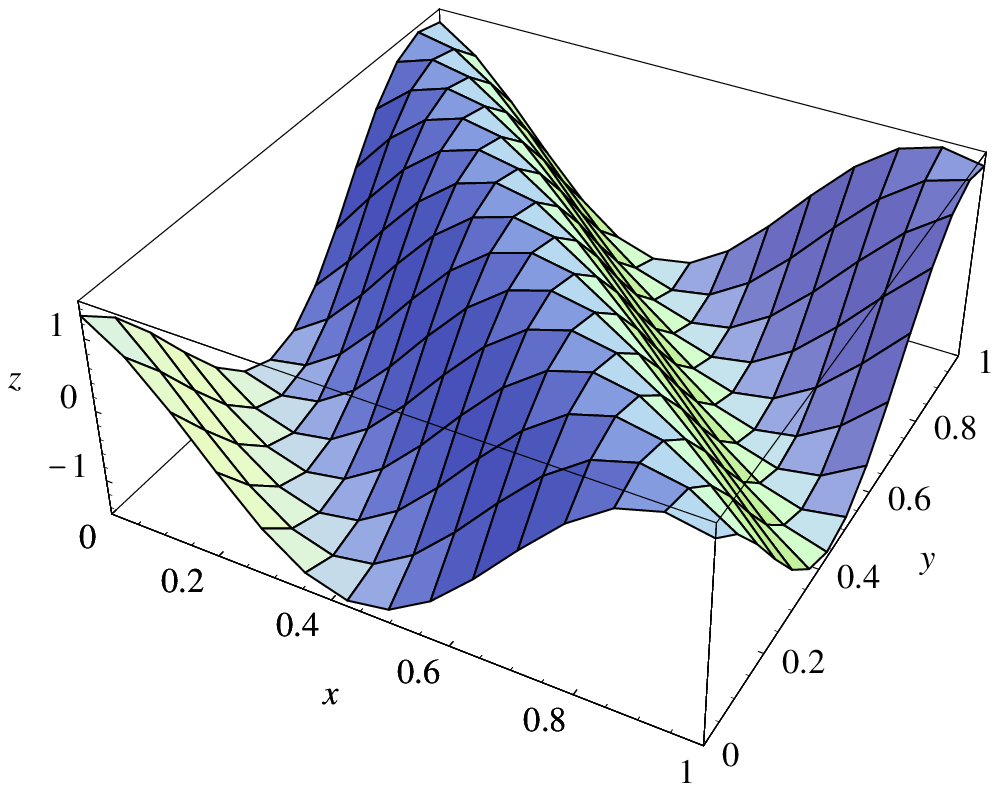}
    \end{center}
    \caption[1]{\small
    \label{wfig}
Representative minimum variance
weightings $w_i(\br)$,
equation~(\protect\ref{wrminvar}),
for the cases
(top) $\bk_i = \{1,0,0\}$,
and
(bottom) $\bk_i = \{1,1,0\}$.
They are just single Fourier modes,
appropriately scaled and phased.
    }
    \end{figure}
}

\newcommand{\tetrahedronfig}{
    \begin{figure}
    \begin{center}
    \leavevmode
    \includegraphics[width=2in]{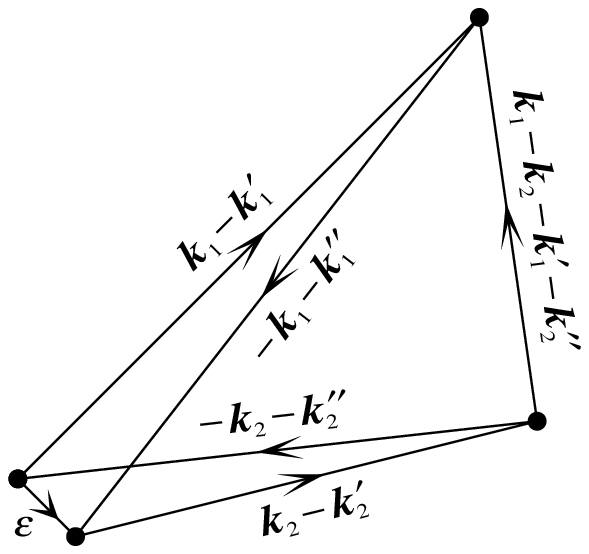}
    \end{center}
    \caption[1]{\small
    \label{tetrahedronfig}
Four-point configuration of wavevectors
for the trispectrum in
equation~(\protect\ref{DphatiDphatj}),
which describes
the covariance of power spectra of weighted densities.
The short leg $\bvarepsilon$,
equation~(\protect\ref{epsilon}),
produces a beat-coupling to large scales.
    }
    \end{figure}
}

\newcommand{\vartwofig}{
    \begin{figure}
    \begin{center}
    \leavevmode
    \includegraphics[width=3.25in]{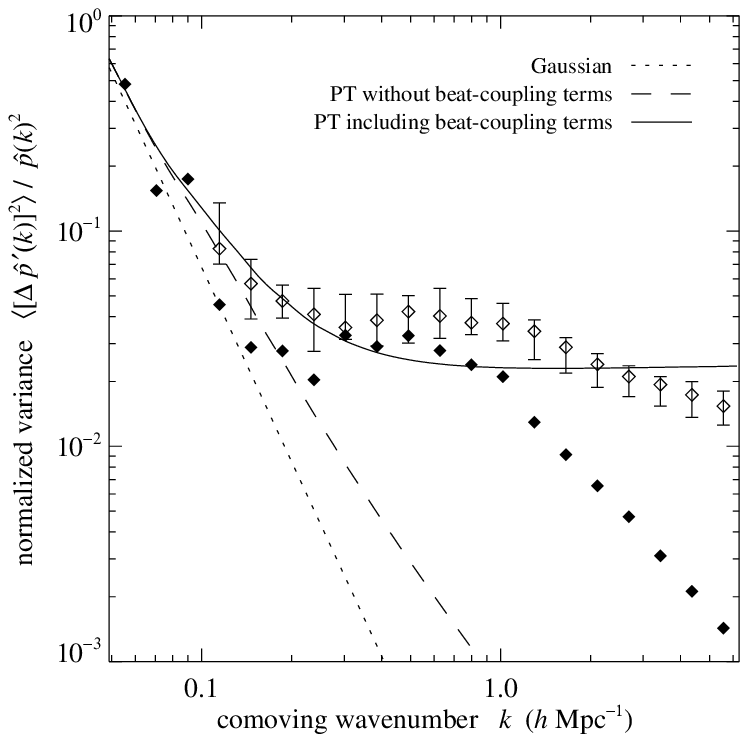}
    \end{center}
    \caption[1]{\small
    \label{vartwofig}
Comparison between the normalized variance
of power measured
from 25 ART $\Lambda$CDM simulations
by
(symbols with error bars, indicating median and quartiles)
the weightings method,
and
(plain symbols)
the ensemble method.
The two methods disagree substantially at nonlinear scales.
Lines show the normalized variance
predicted by perturbation theory
both with (solid line), and without (dashed line)
the large-scale beat-coupling contribution.
The dotted line shows the expected
Gaussian contribution to the variance.
This figure is a condensed version of
Figure~5 of \citet{RH06}.
    }
    \end{figure}
}

\newcommand{\eightpointfig}{
    \begin{figure}
    \begin{center}
    \leavevmode
    \includegraphics[width=2in]{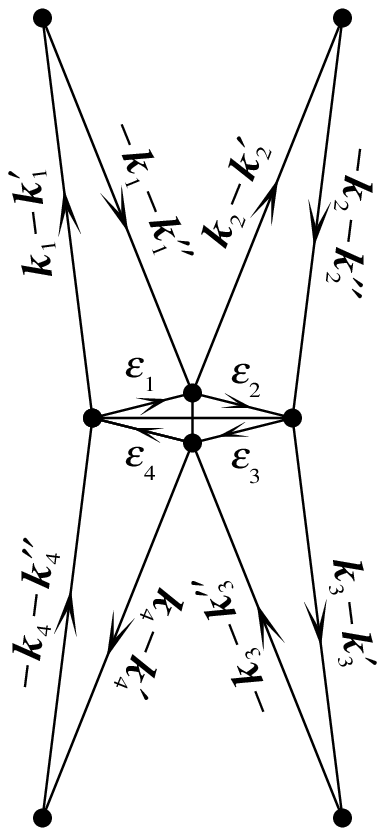}
    \end{center}
    \caption[1]{\small
    \label{eightpointfig}
Eight-point configuration of wavevectors
contributing to the covariance of covariance of power spectra
of weighted densities,
equation~(\protect\ref{Dphati2Dphatj2}).
The central tetrahedron of short legs
produces beat-couplings to large scales.
    }
    \end{figure}
}

\begin{document}

\maketitle

\begin{abstract}
We show how to estimate the covariance of the power spectrum
of a statistically homogeneous and isotropic density field
from a single periodic simulation,
by applying a set of weightings to the density field,
and by measuring the scatter in power spectra between different weightings.
We recommend
a specific set of $52$ weightings
containing only combinations of fundamental modes,
constructed to yield a minimum variance estimate of the covariance of power.
Numerical tests reveal that
at nonlinear scales the variance of power
estimated by the weightings method substantially exceeds
that estimated from a simple ensemble method.
We argue that the discrepancy
is caused by beat-coupling,
in which products of closely spaced Fourier modes
couple by nonlinear gravitational growth to the beat mode between them.
Beat-coupling appears whenever nonlinear power is measured
from Fourier modes with a finite spread of wavevector,
and is therefore present in the weightings method but not the ensemble method.
Beat-coupling inevitably affects real galaxy surveys,
whose Fourier modes have finite width.
Surprisingly,
the beat-coupling contribution dominates the covariance of power
at nonlinear scales,
so that,
counter-intuitively,
it is expected that
the covariance of nonlinear power in galaxy surveys
is dominated not by small scale structure,
but rather by beat-coupling
to the largest scales of the survey.
\end{abstract}



\section{Introduction}
\label{intro}

The last few years have seen
the emergence of a Standard $\Lambda$CDM Model of cosmology
motivated by and consistent with
a wide range of observations, including
the cosmic microwave background,
distant supernovae,
big-bang nucleosynthesis,
large-scale structure,
the abundance of rich galaxy clusters,
and
local measurements of the Hubble constant
(e.g.\ \citealt{Tegmark04b}).

The power spectrum of fluctuations
(of temperature, density, flux, shear, etc.)
is the primary statistic used to constrain cosmological parameters
from observations of the cosmic microwave background
(\citealt{Spergel03}),
of galaxies
(\citealt{Tegmark04a};
\citealt{Cole05};
\citealt{Sanchez05};
\citealt{Eisenstein05}),
of the Lyman alpha forest
(\citealt{Seljak05};
\citealt{LHHHRS05};
\citealt{VH05}),
and of weak gravitational lensing
(\citealt{HYG02};
\citealt{PLWM03};
\citealt{TJ04};
\citealt{Sheldon04}).

From a cosmological standpoint,
the most precious data lie at large, linear scales,
where fluctuations preserve the imprint
of their primordial generation.
A generic, albeit not universal, prediction of inflation
is that primordial fluctuations should be Gaussian.
At large, linear scales,
observations are consistent with fluctuations being Gaussian
\citep{Komatsu03}.

However,
much of the observational data,
especially those involving galaxies,
lies in the translinear or nonlinear regime.
It remains a matter of ongoing research to elucidate
the extent to which nonlinear data can be used to constrain cosmology.

We recently began
\citep{RH05}
a program to measure quantitatively,
from cosmological simulations,
the Fisher information
content of the nonlinear matter power spectrum
(specifically, in the first instance,
the information
about the initial amplitude of the linear power spectrum).
For Gaussian fluctuations,
the power spectrum contains all possible information
about cosmological parameters.
At nonlinear scales,
where fluctuations are non-Gaussian,
it is natural to start by measuring information
in the power spectrum,
although it seems likely
that additional information
resides in the 3-point and higher order correlation functions
(\citealt{TJ04};
\citealt{SS05}).

Measuring the Fisher information in the power spectrum
involves measuring the covariance matrix of power.
For Gaussian fluctuations,
the expected covariance of estimates of power
is known analytically,
but at nonlinear scales
the covariance of power must be estimated from simulations.

A common way to estimate the covariance matrix of a quantity
is to measure its covariance over an ensemble of computer simulations
(\citealt{MW99};
\citealt{SZH99};
\citealt{ZE05}; \citealt{ZDEK05};
\citealt{RH05}).
However,
a reliable estimate of covariance can be computationally expensive,
requiring many, perhaps hundreds
(\citealt{MW99}; \citealt{RH05})
of realizations.
On the other hand it is physically obvious that
the fluctuations in the values of quantities over
the different parts of a single simulation
must somehow encode the covariance of the quantities.
If the covariance could be measured from single simulations,
then it would be possible to measure covariance
from fewer, and from higher quality, simulations.
In any case,
the ability to measure covariance from a single simulation
can be useful in identifying simulations whose statistical
properties are atypical.

A fundamental difficulty with estimating
covariances from single simulations in cosmology
is that the data are correlated over all scales, from small to large.
As described by
\citet{Kunsch89},
such correlations invalidate some of the
``jackknife''
and
``bootstrap''
schemes suggested in the literature.
In jackknife,
variance is inferred
from how much a quantity varies
when some segments of the data are kept, and some deleted.
Bootstrap is like jackknife,
except that deleted segments are replaced with other segments.

As part of the work leading to the present paper,
we investigated a form of the bootstrap procedure,
in which we filled each octant of a simulation cube
with a block of data selected randomly from the cube.
Unfortunately,
the sharp edges of the blocks introduced undesirable small scale power,
which seemed to compromise the effort to measure covariance of power reliably.
Such effects can be mitigated by tapering
\citep{Kunsch89}.
However,
it seemed to us that bootstrapping, like jackknifing,
is a form of re-weighting data,
and that surely the best way to re-weight data
would be to apply the most slowly possible varying weightings.
For a periodic box,
such weightings would be comprised of
the largest scale modes, the fundamentals.

In the present paper, \S\ref{estimate},
we consider applying an arbitrary weighting to
the density of a periodic cosmological simulation,
and we show how the power spectrum
(and its covariance, and the covariance of its covariance)
of the weighted density are related to the true power spectrum
(and its covariance, and the covariance of its covariance).
We confirm mathematically the intuitive idea
that weighting with fundamentals yields the most reliable estimate
of covariance of power.
Multiplying the density in real space by some weighting
is equivalent to convolving the density in Fourier space
with the Fourier transform of the weighting.
This causes the power spectrum
(and its covariance, and the covariance of its covariance)
to be convolved with the Fourier transform of the square
(and fourth, and eighth powers)
of the weighting.
The convolution does least damage when the weighting window is
as narrow as possible in Fourier space,
which means composed of fundamentals.

In \S\ref{weightings}
we show how to design a best set of weightings,
by minimizing the expected variance of
the resulting estimate of covariance of power.
These considerations lead us to recommend a specific
set of $52$ weightings,
each consisting of a combination of fundamental modes.

This paper should have stopped neatly at this point.
Unfortunately, numerical simulations, described in a companion paper
\citep{RH06},
revealed an unexpected (one might say insidious),
substantial discrepancy at nonlinear scales between
the variance of power estimated by the weightings method
and the variance of power estimated by the ensemble method.
In \S\ref{beatcoupling}
we argue that this discrepancy arises from beat-coupling,
a nonlinear gravitational coupling to the large-scale
beat mode between closely spaced nonlinear wavenumbers,
when the power spectrum is measured from Fourier modes
at anything other than infinitely sharp sets of wavenumbers.
Surprisingly,
in cosmologically realistic simulations,
the covariance of power
is dominated at nonlinear scales
by this beat-coupling to large scales.

We discuss the beat-coupling problem in \S\ref{discussion}.
Beat-coupling is relevant to observations
because real galaxy surveys
yield Fourier modes in
finite bands of wavenumber $k$,
of width $\Delta k \sim 1/R$
where $R$ is a chararacteristic linear size of the survey.

Section~\ref{summary} summarizes the results.

\section{Estimating the covariance of power from an ensemble of weighted density fields}
\label{estimate}

The fundamental idea of this paper is
to apply an ensemble of weightings to a
(non-Gaussian, in general) density field,
and to estimate the covariance of the power spectrum
from the scatter in power between different weightings.
This section derives the relation
between the power spectrum of a weighted density field
and the true power spectrum,
along with its expected covariance, and the covariance of its covariance.

It is shown,
equations~(\ref{piapprox}), (\ref{DphatiDphatjf}), and (\ref{Dphati2Dphatj2g}),
that the expected ((covariance of) covariance of)
shell-averaged power of weighted density fields
is simply proportional to the true
((covariance of) covariance of) shell-averaged power,
provided that two approximations are made.
The two approximations are, firstly,
that the power spectrum and trispectrum
are sufficiently slowly varying functions of their arguments,
equations~(\ref{Papprox}) and (\ref{Tapprox}),
and, secondly,
that power is estimated in sufficiently broad shells in $k$-space,
equation~(\ref{broadshellapprox}).
The required approximations are most accurate if the
weightings contain only the largest scale Fourier modes,
such as the weightings containing only fundamental modes
proposed in \S\ref{weightings}.

As will be discussed in \S\ref{beatcoupling},
the apparently innocent assumption, equation~(\ref{Tapprox}),
that the trispectrum is a slowly varying function of its arguments,
is incorrect,
because it sets to zero some important beat-coupling contributions.
However, it is convenient to pretend
in this section and the next, \S\ref{estimate} and \S\ref{weightings},
that the assumption~(\ref{Tapprox}) is true,
and then to consider
in \S\ref{beatcoupling}
how the results are modified
when the beat-coupling contributions to the trispectrum are included.
Ultimately we find,
\S\ref{largescale},
that the weightings method remains valid
when beat-couplings are included,
and,
\S\ref{notquiteweightings},
that the minimum variance weightings derived in \S\ref{weightings},
while no longer exactly minimum variance,
should be close enough to remain good for practical application.

This section is necessarily rather technical,
because it is necessary to distinguish carefully
between various flavours of power spectrum:
estimated versus expected;
unweighted versus weighted;
non-shell-averaged versus shell-averaged.
Subsections~\ref{P} to \ref{ddP}
present expressions for the various power spectra,
their covariances, and the covariances of their covariances.
Subsections~\ref{subtractmeanP}
and \ref{subtractmeanddP}
show how the expressions are modified when,
as is usually the case,
deviations in power must be measured relative to an estimated
rather than an expected value of power.

\subsection{The power spectrum}
\label{P}

Let $\rho(\br)$ denote the density of a statistically homogeneous
random field at position $\br$ in a periodic box.
Choose the unit of length so that the box has unit side.
The density $\rho(\br)$ might represent, perhaps, a realization
of the nonlinearly evolved distribution of dark matter, or of galaxies.
The density could be either continuous or discrete (particles).
Expanded in Fourier modes $\rho(\bk)$,
the density $\rho(\br)$ is\footnote{
The same symbol $\rho$ is used in both real and Fourier space.
The justification for this notation is that $\rho$ is the
same vector in Hilbert space
irrespective of the basis with respect to which it is expanded.
See for example \cite{H05} for a pedagogical exposition.
}
\begin{equation}
\label{rhok}
  \rho(\br) =
  \sum_{\bk} \rho(\bk) \e^{- 2\upi \im \bk . \br}
  \ .
\end{equation}
Thanks to periodicity,
the sum is over an integral lattice of wavenumbers,
$\bk = \{k_x, k_y, k_z\}$
with integer $k_x$, $k_y$, $k_z$.

The expectation value $\langle \rho(\br) \rangle$ of the density
defines the true mean density $\rhobar$,
which without loss of generality we take to equal unity
\begin{equation}
\label{rhobar}
  \rhobar \equiv \langle \rho(\br) \rangle = 1
  \ .
\end{equation}
The deviation $\Delta\rho(\br)$ of the density from the mean is
\begin{equation}
\label{Drhobar}
  \Delta\rho(\br) \equiv \rho(\br) - \rhobar
  \ .
\end{equation}
The expectation values of the Fourier amplitudes vanish,
$\langle \rho(\bk) \rangle = 0$,
except for the zero'th mode,
whose expectation value equals the mean density,
$\langle \rho(\zero) \rangle = \rhobar$.
The Fourier amplitude $\rho(\zero)$ of the zero'th mode is the actual
density of the realization,
which could be equal to, or differ slightly from,
the true mean density $\rhobar$,
depending on whether the mean density of the realization
was constrained to equal the true density, or not.

Because the density field is by assumption statistically homogeneous,
the expected covariance of Fourier amplitudes
$\rho(\bk)$
is a diagonal matrix
\begin{equation}
\label{DrhokDrhok}
  \left\langle \Delta\rho(\bk_1) \, \Delta\rho(\bk_2) \right\rangle
  =
  1_{\bk_1 + \bk_2}
  P(\bk_1)
  \ .
\end{equation}
Here $1_\bk$ denotes the discrete delta-function,
\begin{equation}
  1_\bk =
  \left\{
  \begin{array}{ll}
    1 & \mbox{if $\bk = \zero$} \\
    0 & \mbox{otherwise}
  \end{array}
  \right.
\end{equation}
and $P(\bk)$ is the power spectrum.
Note that there would normally be an extra factor of $\rhobar^{-2}$
on the left hand side of equation~(\ref{DrhokDrhok}),
but it is fine to omit the factor here
because the mean density is normalized to unity,
equation~(\ref{rhobar}).
The reason for dropping the factor of $\rhobar^{-2}$
is to maintain notational consistency with
equation~(\ref{Pi}) below
for the power spectrum of weighted density
(where the deviation in density is necessarily
{\em not\/} divided by the mean).

The symmetry $P(-\bk) = P(\bk)$
in equation~(\ref{DrhokDrhok})
expresses pair exchange symmetry.
Below, \S\ref{shell},
we will assume that the density field is statistical isotropic,
in which case the power is a function $P(k)$
only of the scalar wavenumber $k \equiv |\bk|$,
but for now we stick to the more general case
where power is a function $P(\bk)$ of vector wavenumber $\bk$.

\subsection{The power spectrum of weighted density}
\label{Pw}

Let
$w_i(\br)$
denote the $i$'th member of a set of real-valued weighting functions,
and let $\rho_i(\br)$ denote the density weighted by the
$i$'th weighting
\begin{equation}
  \rho_i(\br) \equiv w_i(\br) \rho(\br)
  \ .
\end{equation}
The Fourier amplitudes $\rho_i(\bk)$
of the weighted density are convolutions
of the Fourier amplitudes of the weighting and the density:
\begin{equation}
\label{rhoik}
  \rho_i(\bk) =
  \sum_{\bk^\prime} w_i(\bk^\prime) \rho(\bk - \bk^\prime)
  \ .
\end{equation}
Reality of the weighting functions implies
\begin{equation}
  w_i(-\bk) = w_i^\ast(\bk)
  \ .
\end{equation}

The expected mean $\rhobar_i(\br)$ of the weighted density
is proportional to the weighting,
\begin{equation}
\label{rhobari}
  \rhobar_i(\br)
  \equiv
  \langle \rho_i(\br) \rangle
  = w_i(\br)
\end{equation}
in which a factor of $\rhobar$ on the right hand side has been omitted
because the mean density has been normalized to unity, equation~(\ref{rhobar}).
The deviation $\Delta\rho_i(\br)$ of the weighted density from the mean is
\begin{equation}
\label{Drhobari}
  \Delta\rho_i(\br) \equiv \rho_i(\br) - \rhobar_i(\br)
  \ .
\end{equation}
In Fourier space the expected mean $\rhobar_i(\bk)$ of the weighted density is
\begin{equation}
\label{rhobarik}
  \rhobar_i(\bk)
  \equiv
  \langle \rho_i(\bk) \rangle
  = w_i(\bk)
\end{equation}
and the deviation $\Delta\rho_i(\bk)$ of the weighted density from the mean is
\begin{equation}
\label{Drhobarik}
  \Delta\rho_i(\bk) \equiv \rho_i(\bk) - \rhobar_i(\bk)
  \ .
\end{equation}
The deviations $\Delta\rho_i(\bk)$
in the Fourier amplitudes of the weighted density are convolutions
of the weighting and the deviation in the density
\begin{equation}
\label{Drhoik}
  \Delta\rho_i(\bk) =
  \sum_{\bk^\prime} w_i(\bk^\prime) \Delta\rho(\bk - \bk^\prime)
\end{equation}
similarly to equation~(\ref{rhoik}).

The expected covariance between
two weighted densities $\rho_i(\bk_1)$ and $\rho_j(\bk_2)$
at wavenumbers $\bk_1$ and $\bk_2$ is,
from equations~(\ref{DrhokDrhok}) and (\ref{Drhoik}),
\begin{equation}
\label{rhoikrhojk}
  \left\langle \Delta\rho_i(\bk_1) \, \Delta\rho_j(\bk_2) \right\rangle
  =
  \sum_{\bk^\prime} w_i(\bk^\prime) w_j(\bk_1 + \bk_2 - \bk^\prime)
  P(\bk_1 - \bk^\prime)
  \ .
\end{equation}
The weighting breaks statistical homogeneity,
so the expected covariance matrix of Fourier amplitudes
$\rho_i(\bk)$,
equation~(\ref{rhoikrhojk}),
is not diagonal.
Nevertheless we {\em define} the power spectrum $P_i(\bk)$
of the $i$'th weighted density
by the diagonal elements of the covariance matrix,
the variance
\begin{equation}
\label{Pi}
  P_i(\bk)
  \equiv
  \left\langle \Delta\rho_i(\bk) \, \Delta\rho_i(- \bk) \right\rangle
  \ .
\end{equation}
Note that this definition~(\ref{Pi}) of the power spectrum $P_i(\bk)$
differs from the usual definition of power in that
the deviations $\Delta\rho_i(\bk)$ on the right
are Fourier transforms of the deviations $\Delta\rho_i(\br)$
{\em not\/} divided by the mean density $\rhobar_i(\br) = w_i(\br)$
(dividing by the mean density would simply unweight the weighting,
defeating the whole point of the procedure).
The power spectrum $P_i(\bk)$ defined by equation~(\ref{Pi})
is related to the true power spectrum $P(\bk)$ by,
equation~(\ref{rhoikrhojk}),
\begin{equation}
  P_i(\bk)
  =
  \sum_{\bk^\prime} \left| w_i(\bk^\prime) \right|^2
  P(\bk - \bk^\prime)
  \ .
\end{equation}

Now make the approximation that the power spectrum
$P(\bk - \bk^\prime)$
at the wavenumber $\bk - \bk^\prime$ displaced by $\bk^\prime$ from $\bk$
is approximately equal to the power spectrum $P(\bk)$
at the undisplaced wavenumber $\bk$
\begin{equation}
\label{Papprox}
  P(\bk - \bk^\prime)
  \approx
  P(\bk)
  \ .
\end{equation}
This approximation is good
provided that the power spectrum $P(\bk)$
is slowly varying as a function of wavenumber $\bk$,
and that the displacement $\bk^\prime$ is small compared to $\bk$.
In \S\ref{weightings} we constrain the weightings $w_i(\bk^\prime)$
to contain only fundamental modes,
$\bk^\prime = \{k^\prime_x, k^\prime_y, k^\prime_z\}$
with $k^\prime_x$, $k^\prime_y$, $k^\prime_z$ = $0, \pm 1$,
so that the displacement $\bk^\prime$ is as small as it can be
without being zero,
and the approximation~(\ref{Papprox})
is therefore as good as it can be.
The approximation~(\ref{Papprox})
becomes exact in the case of a constant, or shot noise,
power spectrum $P(\bk)$,
except at $\bk - \bk^\prime = \zero$.

Under approximation~(\ref{Papprox}),
the power spectrum of the $i$'th weighted density is
\begin{equation}
\label{Piapproxw}
  P_i(\bk)
  \approx
  P(\bk)
  \sum_{\bk^\prime} \left| w_i(\bk^\prime) \right|^2
\end{equation}
which is just proportional to the true power spectrum $P(\bk)$.

Without loss of generality,
let each weighting $w_i(\bk^\prime)$
be normalized so that the factor
on the right hand side of equation~(\ref{Piapproxw}) is unity
\begin{equation}
\label{wnorm}
  \sum_{\bk^\prime} \left| w_i(\bk^\prime) \right|^2
  = 1
  \ .
\end{equation}
Then the power spectrum $P_i(\bk)$
of the weighted density
is approximately equal to the true power spectrum $P(\bk)$
\begin{equation}
\label{Piapprox}
  P_i(\bk)
  \approx
  P(\bk)
  \ .
\end{equation}
Thus, in the approximation~(\ref{Papprox})
and with the normalization~(\ref{wnorm}),
measurements of the power spectrum $P_i(\bk)$ of weighted densities
provide estimates of the true power spectrum $P(\bk)$.
The plan is to use the scatter in the estimates of power
over a set of weightings to estimate the covariance matrix of power.


\subsection{The covariance of power spectra}
\label{cov}

Let $\Phat(\bk)$
denote the power spectrum of unweighted density at wavevector $\bk$
measured from a simulation,
the hat distinguishing it from the true power spectrum $P(\bk)$:
\begin{equation}
\label{Phat}
  \Phat(\bk) \equiv \Delta\rho(\bk) \Delta\rho(-\bk)
  \ .
\end{equation}
Below, \S\ref{shell}, we will invoke statistical isotropy,
and we will average over a shell in $k$-space,
but in equation~(\ref{Phat})
there is no averaging
because there is just one simulation,
and just one specific wavenumber $\bk$.
Because of statistical fluctuations,
the estimate $\Phat(\bk)$ will in general differ from the true power $P(\bk)$,
but by definition
the expectation value of the estimate equals the true value,
$\langle \Phat(\bk) \rangle = P(\bk)$.
The deviation $\Delta\Phat(\bk)$ in the power
is the difference between the measured and expected value:
\begin{equation}
\label{DPhat}
  \Delta\Phat(\bk)
  \equiv
  \Phat(\bk) - P(\bk)
  \ .
\end{equation}

The expected covariance of power
involves the covariance of the covariance of unweighted densities
\begin{eqnarray}
\label{DrhokDrhokDrhokDrhok}
\lefteqn{
  \Bigl\langle
    \bigl[ \Delta\rho(\bk_1) \Delta\rho(\bk_2) - 1_{\bk_1 + \bk_2} P(\bk_1) \bigr]
}
&&
\nonumber
\\
&\times&
    \bigl[ \Delta\rho(\bk_3) \Delta\rho(\bk_4) - 1_{\bk_3 + \bk_4} P(\bk_3) \bigr]
  \Bigr\rangle
\nonumber
\\
&=&
  \left( 1_{\bk_1 + \bk_3} 1_{\bk_2 + \bk_4} + 1_{\bk_1 + \bk_4} 1_{\bk_2 + \bk_3} \right)
  P(\bk_1) P(\bk_2)
\nonumber
\\
&&
  \mbox{}
  +
  1_{\bk_1 + \bk_2 + \bk_3 + \bk_4}
  T(\bk_1, \bk_2, \bk_3, \bk_4)
\label{eta}
\end{eqnarray}
which is a sum of a reducible, Gaussian part,
the terms proportional to $P(\bk_1) P(\bk_2)$,
and an irreducible, non-Gaussian part,
the term involving the trispectrum
$T(\bk_1, \bk_2, \bk_3, \bk_4)$.
Equation~(\ref{eta}) essentially defines what is meant by the trispectrum
$T$.
Exchange symmetry implies that the trispectrum function
is invariant under permutations of its 4 arguments.
The momentum-conserving delta-function $1_{\bk_1 + \bk_2 + \bk_3 + \bk_4}$
in front of the trispectrum $T$ expresses translation invariance.

It follows from equation~(\ref{DrhokDrhokDrhokDrhok}) that
the expected covariance of estimates of power is
\begin{eqnarray}
\label{DPhatDPhat}
\lefteqn{
  \left\langle \Delta\Phat(\bk_1) \Delta\Phat(\bk_2) \right\rangle
}
&&
\\
\nonumber
&=&
  \left( 1_{\bk_1 + \bk_2} + 1_{\bk_1 - \bk_2}\right)
  P(\bk_1)^2
  +
  T(\bk_1, -\bk_1, \bk_2, -\bk_2)
  \ .
\end{eqnarray}

\subsection{The covariance of power spectra of weighted density}
\label{covw}

Similarly to equations~(\ref{Phat}) and (\ref{DPhat}),
let $\Phat_i(\bk)$
denote the power spectrum of the $i$'th weighted density at wavevector $\bk$
measured from a simulation
\begin{equation}
\label{Phati}
  \Phat_i(\bk) \equiv \Delta\rho_i(\bk) \Delta\rho_i(-\bk)
\end{equation}
and let $\Delta\Phat_i(\bk)$ denote the deviation
between the measured and expected value
\begin{equation}
\label{DPhati}
  \Delta\Phat_i(\bk)
  \equiv
  \Phat_i(\bk) - P_i(\bk)
  \ .
\end{equation}
The expected covariance between the power spectra
of the $i$'th and $j$'th weighted densities is,
from equations~(\ref{Drhoik}) and (\ref{DrhokDrhokDrhokDrhok}),
\begin{eqnarray}
\label{DPhatiDPhatj}
\lefteqn{
  \left\langle \Delta\Phat_i(\bk_1) \Delta\Phat_j(\bk_2) \right\rangle
  =
}
&&
\nonumber
\\
\nonumber
\lefteqn{
  \sum_{\bk_1^\prime + \bk_1^{\prime\prime} + \bk_2^\prime + \bk_2^{\prime\prime} = \zero}
  w_i(\bk_1^\prime) w_i(\bk_1^{\prime\prime})
  w_j(\bk_2^\prime) w_j(\bk_2^{\prime\prime})
}
&&
\\
\nonumber
\lefteqn{
  \Bigl[
  \bigl(
    1_{\bk_1 - \bk_1^\prime + \bk_2 - \bk_2^\prime}
    +
    1_{\bk_1 - \bk_1^\prime - \bk_2 - \bk_2^{\prime\prime}}
  \bigr)
  P(\bk_1 {-} \bk_1^\prime) P(- \bk_1 {-} \bk_1^{\prime\prime} )
}
&&
\\
\lefteqn{
  \mbox{}
  +
  T(\bk_1 {-} \bk_1^\prime, -\bk_1 {-} \bk_1^{\prime\prime}, \bk_2 {-} \bk_2^\prime, -\bk_2 {-} \bk_2^{\prime\prime})
  \Bigr]
  \ .
}
&&
\end{eqnarray}

\subsection{The covariance of shell-averaged power spectra}
\label{shell}

Assume now that the unweighted density field $\rhobar(\br)$
is statistically isotropic,
so that the true power spectrum $P(k)$ is a function only of
the absolute value $k \equiv |\bk|$ of its argument.
In estimating the power $P(k)$ from a simulation,
one would typically average the measured power
over a spherical shell $V_k$ of wavenumbers in $k$-space.
Actually the arguments below generalize immediately
to the case where the power is not isotropic,
in which case $V_k$ might be chosen to be some localized patch in $k$-space.
However,
we shall assume isotropy,
and refer to $V_k$ as a shell.

Let $\phat(k)$
denote the measured power averaged over a shell $V_k$
about scalar wavenumber $k$
(the estimated shell-averaged power $\phat(k)$ is written in lower case
to distinguish it from the estimate $\Phat(\bk)$ of power
at a single specific wavevector $\bk$):
\begin{equation}
\label{phat}
  \phat(k) \equiv \frac{1}{N_\bk} \sum_{\bk \in V_k} \Phat(\bk)
  \ .
\end{equation}
Here $N_\bk$ is the number of modes $\rho(\bk)$ in the shell $V_k$.
We count $\rho(\bk)$ and its complex conjugate $\rho(-\bk)$
as contributing two distinct modes,
the real and imaginary parts of $\rho(\bk)$.
The expectation value of the estimates $\phat(k)$
of shell-averaged power equals the true shell-averaged power $p(k)$
\begin{equation}
\label{p}
  \langle \phat(k) \rangle
  = p(k)
  \equiv \frac{1}{N_\bk} \sum_{\bk \in V_k} P(\bk)
  \ .
\end{equation}
The deviation $\Delta\phat(k)$
between the measured and expected value of shell-averaged power is
\begin{equation}
\label{Dphat}
  \Delta\phat(k) \equiv \phat(k) - p(k)
  = \frac{1}{N_\bk} \sum_{\bk \in V_k} \Delta\Phat(\bk)
  \ .
\end{equation}
The expected covariance of shell-averaged estimates of power is,
from equations~(\ref{Dphat}) and (\ref{DPhatDPhat}),
\begin{eqnarray}
\label{DphatDphat}
  \left\langle \Delta\phat(k_1) \Delta\phat(k_2) \right\rangle
&=&
  \frac{1}{N_{\bk_1} N_{\bk_2}}
  \Bigl[
  2
  \sum_{\bk_1 \in V_{k_1} \cap V_{k_2}}
  P(\bk_1)^2
\\
\nonumber
&+&
  \sum_{\bk_1 \in V_{k_1} , \  \bk_2 \in V_{k_2}}
  T(\bk_1, -\bk_1, \bk_2, -\bk_2)
  \Bigr]
  \ .
\end{eqnarray}
In the usual case,
the shells $V_{k}$ would be taken to be non-overlapping,
in which case the intersection
$V_{k_1} \cap V_{k_2}$
in equation~(\ref{DphatDphat})
is equal either to $V_{k_1}$
if $V_{k_1}$ and $V_{k_2}$ are the same shell,
or to the empty set
if $V_{k_1}$ and $V_{k_2}$ are different shells.

\subsection{The covariance of shell-averaged power spectra of weighted density}
\label{shellw}

Similarly to equation~(\ref{phat}),
let $\phat_i(k)$ denote
the measured shell-averaged power spectrum
of the $i$'th weighted density at wavenumber $k$
\begin{equation}
\label{phati}
  \phat_i(k) \equiv \frac{1}{N_\bk} \sum_{\bk \in V_k} \Phat_i(\bk)
  \ .
\end{equation}
The expectation value of the estimates $\phat_i(k)$ is
(compare eq.~(\ref{p}))
\begin{equation}
\label{pi}
  \langle \phat_i(k) \rangle = p_i(k)
  \equiv \frac{1}{N_\bk} \sum_{\bk \in V_k} P_i(\bk)
  \ .
\end{equation}
In the approximation~(\ref{Papprox}) of a slowly varying power spectrum,
and with the normalization~(\ref{wnorm}),
the expected shell-averaged power spectrum $p_i(k)$ of the weighted density
is approximately equal to the
shell-averaged power spectrum $p(k)$ of the unweighted density
(compare eq.~(\ref{Piapprox}))
\begin{equation}
\label{piapprox}
  p_i(k) \approx p(k)
  \ .
\end{equation}
The deviation $\Delta\phat_i(k)$
between the measured and expected values is
(compare eq.~(\ref{Dphat}))
\begin{equation}
\label{Dphati}
  \Delta\phat_i(k)
  \equiv
  \phat_i(k) - p_i(k)
  =
  \frac{1}{N_\bk} \sum_{\bk \in V_k} \Delta\Phat_i(\bk)
  \ .
\end{equation}
The expected covariance of shell-averaged power spectra of weighted densities is,
from equations~(\ref{Dphati}) and (\ref{DPhatiDPhatj}),
\begin{eqnarray}
\label{DphatiDphatj}
\lefteqn{
  \left\langle \Delta\phat_i(k_1) \Delta\phat_j(k_2) \right\rangle
  =
  \frac{1}{N_{\bk_1} N_{\bk_2}}
}
&&
\nonumber
\\
\nonumber
\lefteqn{
  \sum_{\bk_1^\prime + \bk_1^{\prime\prime} + \bk_2^\prime + \bk_2^{\prime\prime} = \zero}
  w_i(\bk_1^\prime) w_i(\bk_1^{\prime\prime})
  w_j(\bk_2^\prime) w_j(\bk_2^{\prime\prime})
  \sum_{\bk_1 \in V_{k_1} , \  \bk_2 \in V_{k_2}}
}
&&
\\
\nonumber
\lefteqn{
  \Bigl[
  \bigl(
    1_{\bk_1 - \bk_1^\prime + \bk_2 - \bk_2^\prime}
    +
    1_{\bk_1 - \bk_1^\prime - \bk_2 - \bk_2^{\prime\prime}}
  \bigr)
  P(\bk_1 {-} \bk_1^\prime) P(- \bk_1 {-} \bk_1^{\prime\prime} )
}
&&
\\
\lefteqn{
  \quad
  \mbox{}
  +
  T(\bk_1 {-} \bk_1^\prime, -\bk_1 {-} \bk_1^{\prime\prime}, \bk_2 {-} \bk_2^\prime, -\bk_2 {-} \bk_2^{\prime\prime})
  \Bigr]
  \ .
}
&&
\end{eqnarray}
Assume, analogously to approximation~(\ref{Papprox}) for the power spectrum,
that the trispectrum function
$T(\bk_1 {-} \bk_1^\prime, -\bk_1 {-} \bk_1^{\prime\prime}, \bk_2 {-} \bk_2^\prime, -\bk_2 {-} \bk_2^{\prime\prime})$
in equation~(\ref{DphatiDphatj})
is sufficiently slowly varying,
and the displacements $\bk_1^\prime$, $\bk_1^{\prime\prime}$, $\bk_2^\prime$, $\bk_2^{\prime\prime}$
sufficiently small,
that
\begin{eqnarray}
\label{Tapprox}
\lefteqn{
  T(\bk_1 {-} \bk_1^\prime, -\bk_1 {-} \bk_1^{\prime\prime}, \bk_2 {-} \bk_2^\prime, -\bk_2 {-} \bk_2^{\prime\prime})
}
\nonumber
\\
&\approx&
  T(\bk_1, - \bk_1, \bk_2, - \bk_2)
  \ .
\end{eqnarray}

In \S\ref{beatcoupling}
we will revisit the approximation~(\ref{Tapprox}),
and show that in fact it is not true,
in a way that proves to be interesting and observationally relevant.
In this section and the next, \S\ref{weightings},
however, we will continue to assume that the approximation~(\ref{Tapprox}) is valid.

In the approximations~(\ref{Papprox}) and (\ref{Tapprox})
that the power spectrum and trispectrum
are both approximately constant for small displacements of their arguments,
the covariance of shell-averaged power spectra,
equation~(\ref{DphatiDphatj}),
becomes
\begin{eqnarray}
\label{DphatiDphatjapprox}
\lefteqn{
  \left\langle \Delta\phat_i(k_1) \Delta\phat_j(k_2) \right\rangle
  \approx
  \frac{1}{N_{\bk_1} N_{\bk_2}}
}
&&
\nonumber
\\
\nonumber
\lefteqn{
  \sum_{\bk_1^\prime + \bk_1^{\prime\prime} + \bk_2^\prime + \bk_2^{\prime\prime} = \zero}
  w_i(\bk_1^\prime) w_i(\bk_1^{\prime\prime})
  w_j(\bk_2^\prime) w_j(\bk_2^{\prime\prime})
}
&&
\\
\nonumber
\lefteqn{
  \sum_{\bk_1 \in V_{k_1} , \  \bk_2 \in V_{k_2}}
  \Bigl[
  \bigl(
    1_{\bk_1 - \bk_1^\prime + \bk_2 - \bk_2^\prime}
    +
    1_{\bk_1 - \bk_1^\prime - \bk_2 - \bk_2^{\prime\prime}}
  \bigr)
  P(\bk_1)^2
}
&&
\\
\lefteqn{
  \qquad\qquad\qquad
  \mbox{}
  +
  T(\bk_1, -\bk_1, \bk_2, -\bk_2)
  \Bigr]
  \ .
}
&&
\end{eqnarray}

Consider the Gaussian ($P^2$) part of this expression~(\ref{DphatiDphatjapprox}).
In the true covariance of shell-averaged power, equation~(\ref{DphatDphat}),
the Gaussian part of the covariance is a diagonal matrix,
with zero covariance between non-overlapping shells.
By contrast,
the Gaussian part of the covariance of power of weighted densities,
equation~(\ref{DphatiDphatjapprox}),
is not quite diagonal.
In effect,
the Gaussian variance in each shell
is smeared by convolution with the weighting function,
causing some of the Gaussian variance
near the boundaries of adjacent shells
to leak into covariance between the shells.
In \S\ref{weightings},
we advocate restricting
the weightings $w_i(\bk)$ to contain only fundamental modes,
which keeps smearing to a minimum.
Whatever the case,
if each shell $V_k$
is broad compared the extent of the weightings $w_i(\bk)$ in $k$-space,
then the smearing is relatively small, and can be approximated as zero.
Mathematically, this broad-shell approximation
amounts to approximating
\begin{eqnarray}
\label{broadshellapprox}
\lefteqn{
  \sum_{\bk_1 \in V_{k_1} , \  \bk_2 \in V_{k_2}}
    1_{\bk_1 - \bk_1^\prime + \bk_2 - \bk_2^\prime}
    +
    1_{\bk_1 - \bk_1^\prime - \bk_2 - \bk_2^{\prime\prime}}
}
&&
\\
\nonumber
&\approx&
  \sum_{\bk_1 \in V_{k_1} , \  \bk_2 \in V_{k_2}}
    1_{\bk_1 + \bk_2}
    +
    1_{\bk_1 - \bk_2}
  =
  \sum_{\bk_1 \in V_{k_1} \cap V_{k_2}}
  2
  \ .
\end{eqnarray}

In the broad-shell approximation~(\ref{broadshellapprox}),
the expected covariance of shell-averaged power
spectra of weighted densities,
equation~(\ref{DphatiDphatjapprox}),
simplifies to
\begin{equation}
\label{DphatiDphatjf}
  \left\langle \Delta\phat_i(k_1) \Delta\phat_j(k_2) \right\rangle
  \approx
  f_{ij}
  \left\langle \Delta\phat(k_1) \Delta\phat(k_2) \right\rangle
\end{equation}
where the factor $f_{ij}$ is
\begin{equation}
\label{fij}
  f_{ij} \equiv
  \sum_{\bk_1^\prime + \bk_1^{\prime\prime} + \bk_2^\prime + \bk_2^{\prime\prime} = \zero}
  w_i(\bk_1^\prime) w_i(\bk_1^{\prime\prime})
  w_j(\bk_2^\prime) w_j(\bk_2^{\prime\prime})
  \ .
\end{equation}
In real (as opposed to Fourier) space, the factor $f_{ij}$ is
\begin{equation}
  f_{ij}
  = \int w_i(\br)^2 w_j(\br)^2 \, \ddd r
  \ .
\end{equation}

Equation~(\ref{DphatiDphatjf}) is the most basic result of the present paper.
It states that the expected covariance between estimates of
power from various weightings is proportional to the
true covariance matrix of power.
The nice thing about the result~(\ref{DphatiDphatjf})
is that the constant of proportionality $f_{ij}$
depends only on the weightings $w_i(\bk)$ and $w_j(\bk)$,
and is independent both of the power spectrum $P(\bk)$
and of the wavenumbers $k_1$ and $k_2$ in the covariance
$\left\langle \Delta\phat_i(k_1) \Delta\phat_j(k_2) \right\rangle$.


\subsection{The covariance of the covariance of shell-averaged power spectra of weighted density}
\label{ddP}

Equation~(\ref{DphatiDphatjf})
provides the formal mathematical justification
for estimating the covariance of power
from the scatter in estimates of power
over an ensemble of weightings of density.
In \S\ref{weightings} we will craft the
weightings $w_i(\bk)$ so as to minimize the expected variance
of the estimated covariance of power.
The resulting weightings are ``best possible'',
within the framework of the technique.
To determine the minimum variance estimator,
it is necessary to have an expression for
the (co)variance of the covariance of power,
which we now derive.

The expected covariance between estimates
$\Delta\phat_i(k_1) \Delta\phat_i(k_2)$
of covariance of power
is a covariance of covariance of covariance of densities,
an 8-point object.
This object involves, in addition to the 8-point function,
a linear combination of products of lower-order functions adding to 8 points.
The types of terms are
(cf.\ \citealt{VH01})
\begin{equation}
\label{eightpt}
  2^4, \quad 2 \cdot 3^2, \quad 2^2 \cdot 4, \quad 2 \cdot 6, \quad 3 \cdot 5, \quad 4^2, \quad 8
\end{equation}
in which $2^4$ signifies a product of four 2-point functions,
$2 \cdot 3^2$ signifies a product of a 2-point function with two 3-point functions,
and so on, up to $8$, which signifies the 8-point function.
We do not pause to write out all the terms explicitly,
because in the same slowly-varying and broad-shell approximations
that led to equation~(\ref{DphatiDphatjf}),
the covariance of covariance of power spectra of weighted densities simplifies to
\begin{eqnarray}
\label{Dphati2Dphatj2g}
\lefteqn{
  \Bigl\langle
    \bigl[ \Delta\phat_i(k_1) \Delta\phat_i(k_2) - \left\langle \Delta\phat_i(k_1) \Delta\phat_i(k_2) \right\rangle \bigr]
}
&&
\nonumber
\\
&&
    \times
    \bigl[ \Delta\phat_j(k_3) \Delta\phat_j(k_4) - \left\langle \Delta\phat_j(k_3) \Delta\phat_j(k_4) \right\rangle \bigr]
  \Bigr\rangle
\nonumber
\\
&\approx&
  g_{ij}
  \Bigl\langle
    \bigl[ \Delta\phat(k_1) \Delta\phat(k_2) - \left\langle \Delta\phat(k_1) \Delta\phat(k_2) \right\rangle \bigr]
\nonumber
\\
&&
    \times
    \bigl[ \Delta\phat(k_3) \Delta\phat(k_4) - \left\langle \Delta\phat(k_3) \Delta\phat(k_4) \right\rangle \bigr]
  \Bigr\rangle
\end{eqnarray}
where $g_{ij}$ is, analogously to equation~(\ref{fij}),
\begin{eqnarray}
\label{gij}
\lefteqn{
  g_{ij}
  \equiv
  \sum_{\bk_1^\prime + \bk_1^{\prime\prime} + \bk_2^\prime + \bk_2^{\prime\prime} + \bk_3^\prime + \bk_3^{\prime\prime} + \bk_4^\prime + \bk_4^{\prime\prime} = \zero}
}
&&
\\
\nonumber
\lefteqn{
  \quad
  w_i(\bk_1^\prime) w_i(\bk_1^{\prime\prime}) w_i(\bk_2^\prime) w_i(\bk_2^{\prime\prime})
  w_j(\bk_3^\prime) w_j(\bk_3^{\prime\prime}) w_j(\bk_4^\prime) w_j(\bk_4^{\prime\prime})
  \ .
}
&&
\end{eqnarray}
In real (as opposed to Fourier) space, the factors $g_{ij}$ are
\begin{equation}
  g_{ij}
  = \int w_i(\br)^4 w_j(\br)^4 \, \ddd r
  \ .
\end{equation}

Equation~(\ref{Dphati2Dphatj2g}) states,
analogously to equation~(\ref{DphatiDphatjf}),
that the expected covariance of covariance of power spectra
of weighted densities is proportional to the
true covariance of covariance of power.
As with the factors $f_{ij}$, equation~(\ref{fij}),
the constants of proportionality $g_{ij}$, equation~(\ref{gij}),
depend only on the weightings $w_i(\bk)$ and $w_j(\bk)$,
and are independent of the power spectrum $P(\bk)$
or of any of the higher order functions,
and are also independent of the wavenumbers $k_1$, ..., $k_4$
in the covariance,
a gratifyingly simple result.

\subsection{Subtracting the mean power}
\label{subtractmeanP}

The deviation $\Delta\phat_i(k)$
of the shell-averaged power spectrum of the $i$'th weighted density
was defined above, equation~(\ref{Dphati}),
to be the difference between the measured value
$\phat_i(k)$
and the expected value
$p_i(k)$
of shell-averaged power.
However, the expected power spectrum
$p_i(k)$
(the true power spectrum)
is probably unknown.
Even if the true power spectrum is known in the linear regime
(because the simulation was set up with a known linear power spectrum),
the true power spectrum in the non-linear regime is not known precisely,
but must be estimated from the simulation.

In practice, therefore, it is necessary to measure the deviation in power
not from the true value,
but rather from some estimated mean value.
Two strategies naturally present themselves.
The first strategy is to take the mean power spectrum
to be the measured power spectrum
$\phat(k)$
of the unweighted density of the simulation.
In this case the deviation
$\Delta\phat^\prime_i(k)$
between the measured shell-averaged power spectra
of the weighted and unweighted densities is
(the deviation $\Delta\phat^\prime_i(k)$
is primed to distinguish it
from the deviation $\Delta\phat_i(k)$,
eq.~(\ref{Dphati}))
\begin{equation}
\label{Dphati1}
  \Delta\phat^\prime_i(k)
  \equiv
  \phat_i(k) - \phat(k)
  \ .
\end{equation}
The second strategy is to take the mean power spectrum
to be the average over weightings of the measured power spectra
of weighted densities,
$N^{-1} \sum_i \phat_i(k)$.
In this case the deviation
$\Delta\phat^\prime_i(k)$
between the measured shell-averaged power spectra
and their average is
(with the same primed notation for the deviation
$\Delta\phat^\prime_i(k)$
as in eq.~(\ref{Dphati1});
it is up to the user to decide which strategy to adopt)
\begin{equation}
\label{Dphati2}
  \Delta\phat^\prime_i(k)
  \equiv
  \phat_i(k) - \frac{1}{N} \sum_i \phat_i(k)
  \ .
\end{equation}
The advantage of the first strategy,
equation~(\ref{Dphati1}),
is that the power spectrum $\phat(k)$ of
the unweighted density is the most accurate
(by symmetry)
estimate of the power spectrum
that can be measured from a single simulation.
Its disadvantage is that measurements of power spectra of weighted densities
yield (slightly) biassed estimates of the power spectrum of unweighted density,
because the approximation~(\ref{Papprox}) can lead to a slight bias
if, as is typical, the power spectrum $P(\bk)$ is not constant.
In other words, the approximation $p_i(k) \approx p(k)$,
equation~(\ref{piapprox}), is not an exact equality.
Although the bias is likely to be small, it contributes systematically
to estimates of deviations of power,
causing the covariance of power to be systematically over-estimated.
The second strategy,
equation~(\ref{Dphati2}),
is unaffected by this bias,
but the statistical uncertainty is slightly larger.
Probably the sensible thing to do is to apply both strategies,
and to check that they yield consistent results.

To allow a concise expression for the covariance of power to be written down,
it is convenient to introduce $v_i(\bk)$,
defined to be the Fourier transform of the squared real-space weighting,
$v_i(\br) \equiv w_i(\br)^2$,
\begin{equation}
\label{vi}
  v_i(\bk) \equiv \sum_{\bk^\prime + \bk^{\prime\prime} = \bk}
  w_i(\bk^\prime) w_i(\bk^{\prime\prime})
  \ .
\end{equation}
The normalization condition~(\ref{wnorm}) on the weightings $w_i(\bk)$
is equivalent to requiring
\begin{equation}
  v_i(\zero) = 1
  \ .
\end{equation}
In terms of $v_i(\bk)$,
the factors $f_{ij}$,
equation~(\ref{fij}),
relating the expected covariance matrix of power spectra of weighted densities
to the true covariance matrix of power are
\begin{equation}
\label{fijv}
  f_{ij}
  =
  \sum_\bk v_i(\bk) v_j(-\bk)
  \ .
\end{equation}

An expression is desired for the covariance of power
in terms of the deviations $\Delta\phat^\prime_i(k)$,
equations~(\ref{Dphati1}) or (\ref{Dphati2}),
instead of $\Delta\phat_i(k)$.
For this, a modified version of $v_i(\bk)$ is required.
For strategy one,
equation~(\ref{Dphati1}),
\begin{equation}
\label{vpi1}
  v^\prime_i(\bk)
  =
  \left\{
  \begin{array}{ll}
    0 & (\bk = \zero) \\
    v_i(\bk) & (\bk \neq \zero)
  \end{array}
  \right.
\end{equation}
whereas for strategy two,
equation~(\ref{Dphati2}),
\begin{equation}
\label{vpi2}
  v^\prime_i(\bk)
  =
  v_i(\bk) - \frac{1}{N} \sum_i v_i(\bk)
  \ .
\end{equation}
In either case,
the expected covariance
$\left\langle \Delta\phat^\prime_i(k_1) \Delta\phat^\prime_j(k_2) \right\rangle$
of estimates of shell-averaged power spectra is
related to the true covariance
$\left\langle \Delta\phat(k_1) \Delta\phat(k_2) \right\rangle$
of shell-averaged power by
(compare eq.~(\ref{DphatiDphatjf}))
\begin{equation}
\label{DphatiDphatjfp}
  \left\langle \Delta\phat^\prime_i(k_1) \Delta\phat^\prime_j(k_2) \right\rangle
  \approx
  f^\prime_{ij}
  \left\langle \Delta\phat(k_1) \Delta\phat(k_2) \right\rangle
\end{equation}
where the factors $f^\prime_{ij}$ are
(compare eq.~(\ref{fijv}))
\begin{equation}
\label{fpijv}
  f^\prime_{ij}
  =
  \sum_\bk v^\prime_i(\bk) v^\prime_j(-\bk)
  \ .
\end{equation}
The approximation~(\ref{DphatiDphatjfp})
is valid under the same assumptions made
in deriving the approximation~(\ref{DphatiDphatjf}),
namely the slowly-varying approximations~(\ref{Papprox}) and (\ref{Tapprox}),
and the broad-shell approximation~(\ref{broadshellapprox}).

\subsection{Subtracting the mean covariance of power}
\label{subtractmeanddP}

The expression~(\ref{gij})
for the covariance of covariance of power
must likewise be modified
to allow for the fact that the deviations in power must be measured
as deviations not from the true power spectrum but from
either (strategy~1) the power spectrum of the unweighted density,
or (strategy~2) the averaged power spectrum of the weighted densities.

For this purpose it is convenient to define $u_i(\bk)$
to be the Fourier transform of the fourth power of the real-space weighting,
$u_i(\br) \equiv v_i(\br)^2 = w_i(\br)^4$,
\begin{equation}
\label{ui}
  u_i(\bk) \equiv \sum_{\bk^\prime + \bk^{\prime\prime} = \bk}
  v_i(\bk^\prime) v_i(\bk^{\prime\prime})
  \ .
\end{equation}
In terms of $u_i(\bk)$,
the factors $g_{ij}$,
equation~(\ref{gij}),
relating the expected covariance of covariance of power spectra of weighted densities
to the true covariance of covariance of power are
\begin{equation}
\label{giju}
  g_{ij}
  =
  \sum_\bk u_i(\bk) u_j(-\bk)
  \ .
\end{equation}

To write down an expression for
the covariance of the covariance of the deviations
$\Delta\phat^\prime_i(k)$
instead of
$\Delta\phat_i(k)$,
define a modified version $u^\prime_i(\bk)$
of $u_i(\bk)$
by
\begin{equation}
\label{upi}
  u^\prime_i(\bk) \equiv \sum_{\bk^\prime + \bk^{\prime\prime} = \bk}
  v^\prime_i(\bk^\prime) v^\prime_i(\bk^{\prime\prime})
\end{equation}
which is the same as equation~(\ref{ui})
but with primed $v^\prime_i(\bk)$,
equations~(\ref{vpi1}) or (\ref{vpi2}),
in place of $v_i(\bk)$.
Then the covariance of the covariance of the deviations
$\Delta\phat^\prime_i(k)$
is related to the true covariance of covariance of shell-averaged power by
(compare eq.~(\ref{Dphati2Dphatj2g}))
\begin{eqnarray}
\label{Dphati2Dphatj2gp}
\lefteqn{
  \Bigl\langle
    \bigl[ \Delta\phat^\prime_i(k_1) \Delta\phat^\prime_i(k_2) - \left\langle \Delta\phat^\prime_i(k_1) \Delta\phat^\prime_i(k_2) \right\rangle \bigr]
}
&&
\nonumber
\\
&&
    \times
    \bigl[ \Delta\phat^\prime_j(k_3) \Delta\phat^\prime_j(k_4) - \left\langle \Delta\phat^\prime_j(k_3) \Delta\phat^\prime_j(k_4) \right\rangle \bigr]
  \Bigr\rangle
\nonumber
\\
&\approx&
  g^\prime_{ij}
  \Bigl\langle
    \bigl[ \Delta\phat(k_1) \Delta\phat(k_2) - \left\langle \Delta\phat(k_1) \Delta\phat(k_2) \right\rangle \bigr]
\nonumber
\\
&&
    \times
    \bigl[ \Delta\phat(k_3) \Delta\phat(k_4) - \left\langle \Delta\phat(k_3) \Delta\phat(k_4) \right\rangle \bigr]
  \Bigr\rangle
\end{eqnarray}
where the factors $g^\prime_{ij}$ are
(compare eq.~(\ref{giju}))
\begin{equation}
\label{gpiju}
  g^\prime_{ij}
  =
  \sum_\bk u^\prime_i(\bk) u^\prime_j(-\bk)
  \ .
\end{equation}

Equation~(\ref{Dphati2Dphatj2gp})
gives the expected covariance of the difference between
the estimate of covariance
$\Delta\phat^\prime_i(k_1) \Delta\phat^\prime_i(k_2)$
and its expectation value
$\left\langle \Delta\phat^\prime_i(k_1) \Delta\phat^\prime_i(k_2) \right\rangle$,
but this latter expectation value is again an unknown quantity.
What can actually be measured is the difference between
the estimate
$\Delta\phat^\prime_i(k_1) \Delta\phat^\prime_i(k_2)$
and its average over weightings
$N^{-1} \sum_i \Delta\phat^\prime_i(k_1) \Delta\phat^\prime_i(k_2)$.
To write down an expression for the covariance of the covariance
relative to the weightings-averaged covariance
rather than the expected covariance,
define a modified version $u^{\prime\prime}_i(\bk)$ of $u^\prime_i(\bk)$,
equation~(\ref{upi}), by
\begin{equation}
\label{uppi}
  u^{\prime\prime}_i(\bk)
  \equiv
  u^\prime_i(\bk) - \frac{1}{N} \sum_i u^\prime_i(\bk)
  \ .
\end{equation}
Then the covariance of the covariance of the deviations
$\Delta\phat^\prime_i(k)$
is related to the true covariance of covariance of shell-averaged power by
(compare eqs.~(\ref{Dphati2Dphatj2g}) and (\ref{Dphati2Dphatj2gp}))
\begin{eqnarray}
\label{Dphati2Dphatj2gpp}
\lefteqn{
  \Bigl\langle
    \bigl[ \Delta\phat^\prime_i(k_1) \Delta\phat^\prime_i(k_2) - \frac{1}{N} \sum_k \Delta\phat^\prime_k(k_1) \Delta\phat^\prime_k(k_2) \bigr]
}
&&
\nonumber
\\
&&
    \times
    \bigl[ \Delta\phat^\prime_j(k_3) \Delta\phat^\prime_j(k_4) - \frac{1}{N} \sum_l \Delta\phat^\prime_l(k_3) \Delta\phat^\prime_l(k_4) \bigr]
  \Bigr\rangle
\nonumber
\\
&\approx&
  g^{\prime\prime}_{ij}
  \Bigl\langle
    \bigl[ \Delta\phat(k_1) \Delta\phat(k_2) - \left\langle \Delta\phat(k_1) \Delta\phat(k_2) \right\rangle \bigr]
\nonumber
\\
&&
    \times
    \bigl[ \Delta\phat(k_3) \Delta\phat(k_4) - \left\langle \Delta\phat(k_3) \Delta\phat(k_4) \right\rangle \bigr]
  \Bigr\rangle
\end{eqnarray}
where the factors $g^{\prime\prime}_{ij}$ are
(compare eqs.~(\ref{giju}) and (\ref{gpiju}))
\begin{equation}
\label{gppiju}
  g^{\prime\prime}_{ij}
  =
  \sum_\bk u^{\prime\prime}_i(\bk) u^{\prime\prime}_j(-\bk)
  \ .
\end{equation}
Approximations~(\ref{Dphati2Dphatj2gp}) and (\ref{Dphati2Dphatj2gpp})
are valid under the same approximations as
approximations~(\ref{DphatiDphatjf}) and (\ref{Dphati2Dphatj2g}),
namely
the slowly-varying approximations~(\ref{Papprox}) and (\ref{Tapprox}),
and the broad-shell approximation~(\ref{broadshellapprox}).

\section{Minimum variance weightings}
\label{weightings}

It was shown in \S\ref{estimate}
that the expected covariance between shell-averaged power spectra
of weighted densities is proportional to the true covariance of
shell-average power, equation~(\ref{DphatiDphatjfp}).
It follows that the scatter in estimates of power from different weightings
can be used to estimate the true covariance of power.
In this section we use minimum variance arguments
to derive a set of $52$ weightings,
equation~(\ref{wkminvar}),
which we recommend, \S\ref{recommend}, for practical application.

In this section as in the previous one, \S\ref{estimate},
we continue to ignore the beat-coupling contributions
to the (covariance of) covariance of power.
These beat-couplings are discussed in \S\ref{beatcoupling},
which in \S\ref{notquiteweightings} concludes
that the minimum variance weightings derived in the present section,
although no longer precisely minimum variance,
should be satisfactory for practical use.

\subsection{Fundamentals and symmetries}
\label{fund}

In the first place,
we choose to use weightings $w_i(\bk)$
that contain only combinations of fundamental modes,
that is,
$\bk = \{k_x, k_y, k_z\}$
with $k_x$, $k_y$, $k_z$ running over $0, \pm 1$.
By restricting the weightings to fundamental modes only,
we ensure that the two approximations
required for equation~(\ref{DphatiDphatjfp}) to be valid
are as good as can be.
The first approximation was
the slowly-varying approximation,
that both the power spectrum $P(\bk)$
and the trispectrum $T(\bk_1, -\bk_1, \bk_2, -\bk_2)$
remain approximately constant, equations~(\ref{Papprox}) and (\ref{Tapprox}),
when their arguments are displaced by the extent of the weightings $w_i(\bk^\prime)$,
that is,
by amounts $\bk^\prime$ for which $w_i(\bk^\prime)$ is non-zero.
The second approximation was
the broad-shell approximation,
that the shells $V_k$ over which the estimated power $\phat_i(k)$ is averaged
are broad compared to the extent of the weightings $w_i(\bk^\prime)$,
which reduces the relative importance of smearing of Gaussian variance
from the edges of adjacent shells into covariance between the shells.

In the second place,
we choose to use weightings that are symmetrically related to each other,
which seems a natural thing to do given the cubic symmetry of a periodic box.
Choosing a symmetrically related set of weightings
not only simplifies practical application of the procedure,
but also simplifies the mathematics of determining a best set
of Fourier coefficients $w_i(\bk)$,
as will be seen in \S\ref{minvar} below.

There are $48$ rotational and reflectional transformations of a cube,
corresponding to choosing the $x$-axis in any of 6 directions,
then the $y$-axis in any of 4 directions perpendicular to the $x$-axis,
and finally the $z$-axis in either of the 2 directions
perpendicular to the $x$- and $y$-axes.

To the rotational and reflectional transformations
we adjoin the possibility of translations
by a fraction (half, quarter, eighth) of a box along any of the 3 axes,
for a net total of
$48 \times 8^3 = 24{,}576$
possible transformations.
In practice, however,
the minimum variance weightings $w_i(\bk)$
presented in \S\ref{theweightings}
prove to possess a high degree of symmetry,
greatly reducing the number of
distinct weightings.


\subsection{How to derive minimum variance weightings}
\label{minvar}

For brevity,
let $\Xhat_i$ denote an estimate of the covariance of shell-averaged power
from the $i$'th weighted density
(the arguments $k_1$ and $k_2$ on $\Xhat_i$ are suppressed,
since they play no role in the arguments that follow)
\begin{equation}
\label{Xhati}
  \Xhat_i
  \equiv
  \frac{1}{f^\prime} \,
  \Delta\phat^\prime_i(k_1) \Delta\phat^\prime_i(k_2)
  \ .
\end{equation}
The quantity $f^\prime$ here is any diagonal element
\begin{equation}
  f^\prime \equiv f^\prime_{ii}
\end{equation}
of the matrix of factors $f^\prime_{ij}$
defined by equation~(\ref{fpijv});
the diagonal elements $f^\prime_{ii}$ are identically equal for all $i$
because the weightings $w_i(\bk)$ are by assumption symmetrically related.
The factor $1/f^\prime$ in equation~(\ref{Xhati})
ensures that
$\Xhat_i$
is,
in accordance with equation~(\ref{DphatiDphatjfp}),
an
estimate
of the true covariance of shell-averaged power,
which we abbreviate $X$,
\begin{equation}
\label{Xi}
  \langle \Xhat_i \rangle
  \approx X
  \equiv
  \left\langle \Delta\phat(k_1) \Delta\phat(k_2) \right\rangle
  \ .
\end{equation}
The approximation~(\ref{Xi}) is valid
under the assumptions made in deriving equation~(\ref{DphatiDphatjfp}),
namely the slowly-varying approximations~(\ref{Papprox}) and (\ref{Tapprox}),
and the broad-shell approximation~(\ref{broadshellapprox}).

Let $N$ denote the number of weightings.
Because the weightings are by assumption symmetrically related,
it follows immediately that the best estimate
of the true covariance of shell-averaged power
$\left\langle \Delta\phat(k_1) \Delta\phat(k_2) \right\rangle$
will be a straight average over the ensemble of weightings
\begin{equation}
\label{Xhat}
  \Xhat
  =
  \frac{1}{N} \sum_{i} \Xhat_i
  \ .
\end{equation}

It remains to determine the best Fourier coefficients $w_i(\bk)$
for a representative weighting $i$.
The best set is that which minimizes the expected variance
$\langle \Delta\Xhat^2 \rangle \equiv \langle (\Xhat{-}X)^2 \rangle$
of the estimate~(\ref{Xhat}).
According to equation~(\ref{Dphati2Dphatj2gp}),
this expected variance
$\langle \Delta\Xhat^2 \rangle$
is approximately proportional to a factor that depends on the weightings
\begin{equation}
\label{DXhat2}
  \langle \Delta \Xhat^2 \rangle
  =
  \frac{1}{N^2} \sum_{ij}
  \langle \Delta \Xhat_i \Delta \Xhat_j \rangle
  \simpropto
  \frac{1}{(f^\prime N)^2} \sum_{ij} g^\prime_{ij}
\end{equation}
multiplied by another factor that is independent of weightings,
namely the true covariance of covariance of power,
the expression to the right of the coefficient $g^\prime_{ij}$
in equation~(\ref{Dphati2Dphatj2gp}).
Note that the variance $\langle \Delta \Xhat^2 \rangle$
is the expected variance
$\langle (\Xhat{-}X)^2 \rangle$
about the true value $X$,
so it is $g^\prime_{ij}$, equation~(\ref{gpiju}),
not $g^{\prime\prime}_{ij}$, equation~(\ref{gppiju}),
that appears in equation~(\ref{DXhat2}).

Equation~(\ref{DXhat2})
shows that minimizing the variance
$\langle \Delta\Xhat^2 \rangle$
with respect to the coefficients $w_i(\bk)$ of the weightings
is equivalent to minimizing the quantity on the right hand side
of the proportionality~(\ref{DXhat2}).
From equations~(\ref{fpijv}), (\ref{upi}), and (\ref{gpiju})
it follows that this factor can be written
\begin{equation}
\label{DXhat2u}
  \frac{1}{(f^\prime N)^2} \sum_{ij} g^\prime_{ij}
  =
  \frac{1}{u^\prime(\zero)^2}
  \sum_{\bk}
  \left| u^\prime(\bk) \right|^2
\end{equation}
where $u^\prime(\bk)$ denotes the average of $u^\prime_i(\bk)$,
equation~(\ref{upi}),
over weightings
\begin{equation}
\label{up}
  u^\prime(\bk)
  \equiv
  \frac{1}{N}\sum_{i} u^\prime_i(\bk)
  \ .
\end{equation}
Note that $f^\prime = u^\prime(\zero)$.
Equation~(\ref{DXhat2u}) shows that minimizing the variance
$\langle \Delta\Xhat^2 \rangle$
involves computing $u^\prime(\bk)$, equation~(\ref{up}).
We evaluate
$u^\prime(\bk)$
using an algebraic manipulation program (Mathematica) as follows.

A representative weighting $w_i(\bk)$
contains $27$ non-zero Fourier coefficients,
since by assumption it contains only combinations of fundamental modes.
The coefficients $w_i(\bk)$ and $w_i(-\bk)$,
which are complex conjugates of each other,
effectively contribute two coefficients, the real and imaginary parts of $w_i(\bk)$.

First,
evaluate $v_i(\bk)$, equation~(\ref{vi}), in terms of the coefficients $w_i(\bk)$
of the representative weighting.
The $v_i(\bk)$ are non-zero for 125 values of $\bk$,
those whose components
$k_x$, $k_y$, $k_z$ run over $0, \pm 1, \pm 2$.
Each $v_i(\bk)$ is a quadratic polynomial in the $27$ Fourier coefficients.

Next, modify $v_i(\bk)$ to get $v^\prime_i(\bk)$, equation~(\ref{vpi1}),
by setting the coefficient for $\bk = \zero$ to zero.
Again,
each $v^\prime_i(\bk)$ is a quadratic polynomial in the $27$ Fourier coefficients.
For definiteness, we adopt strategy one, equation~(\ref{vpi1}),
rather than strategy two, equation~(\ref{vpi2}).
That is, we assume that the deviation $\Delta\phat^\prime_i(k)$
in the power spectrum of the $i$'th weighting of density
is being measured relative to the power spectrum of the unweighted density,
rather than relative to the average of the power spectra of the weighted densities.
In the end it turns out, \S\ref{theweightings2},
that the minimum variance solution is the same for both strategies,
so there is no loss in restricting to strategy one.

Next, evaluate $u^\prime_i(\bk)$, equation~(\ref{upi}).
The $u^\prime_i(\bk)$ are non-zero for 729 values of $\bk$,
those whose components
$k_x$, $k_y$, $k_z$ run over $0, \pm 1, \mbox{...}, \pm 4$.
Each $u^\prime_i(\bk)$ is a quartic polynomial in the $27$ Fourier coefficients.

Next, evaluate $u^\prime(\bk)$, equation~(\ref{up}),
the average of $u^\prime_i(\bk)$ over weightings $i$.
Consider first averaging $u^\prime_i(\bk)$ over the
48 different rotational and reflectional transformations
of the weighting.
The averaged result $u^\prime(\bk)$
possesses rotational and reflectional symmetry,
so that $u^\prime(\bk)$
is equal to its value at $\bk$
with components permuted and reflected
in such a way that
$0 \leq k_x \leq k_y \leq k_z \leq 4$,
of which there are $35$ distinct cases.
The rotationally and translationally symmetrized function $u^\prime(\bk)$
can be computed by averaging the values of $u^\prime_i(\bk)$
at $729$ values of $\bk$
into $35$ distinct bins.

The symmetrized function satisfies $u^\prime(\bk) = u^\prime(-\bk)$,
so is necessarily real.
Thus the absolute value sign around $u^\prime(\bk)^2$
in equation~(\ref{DXhat2u})
can be omitted.

Now consider averaging the $u^\prime_i(\bk)$ over
translations by half a box in each dimension.
There are $2^3 = 8$ such translations,
and each translation is characterized by a triple
$s_x$, $s_y$, $s_z$
giving the number of half boxes translated in each dimension,
either zero or one for each component.
The effect of the translation is to multiply each
coefficient $w_i(\bk)$
by $(-)^{s_x k_x + s_y k_y + s_z k_z}$,
that is,
by $\pm 1$ according to whether
$s_x k_x + s_y k_y + s_z k_z$ is even or odd.
The sign change carries through the definitions~(\ref{vi}) of $v_i(\bk)$
and (\ref{vpi1}) of $v^\prime_i(\bk)$
to the definition~(\ref{upi}) of $u^\prime_i(\bk)$,
and thence to the definition~(\ref{up}) of $u^\prime(\bk)$.
That is,
the effect of a translation by half a box is to multiply
$u^\prime(\bk)$
by $(-)^{s_x k_x + s_y k_y + s_z k_z}$.
It follows that, after averaging over translations,
$u^\prime(\bk)$ vanishes if any component of $\bk$ is odd,
leaving only cases where all components of $\bk$ are even.
Consequently,
$u^\prime_i(\bk)$ need be evaluated only at the $125$ wavevectors $\bk$
all of whose components are even.
The symmetrized function $u^\prime(\bk)$
can be computed by averaging the values of $u^\prime_i(\bk)$
at the $125$ values of $\bk$
into the $10$ distinct bins with even
$0 \leq k_x \leq k_y \leq k_z \leq 4$.
It is amusing that increasing the number of weightings
(by a factor $8$, if all translations yield distinct weightings)
actually decreases the computational work
required to find the best Fourier coefficients $w_i(\bk)$.

Adjoining translations by a quarter of a box
simplifies the problem of finding the minimum variance solution
for the coefficients $w_i(\bk)$ even further.
There are $4^3 = 64$ such translations,
and each translation is characterized by a triple
$s_x$, $s_y$, $s_z$, each component running over $0$ to $3$,
giving the number of quarter boxes translated in each dimension.
The effect of the translation is to multiply each
coefficient $w_i(\bk)$
by $\im^{s_x k_x + s_y k_y + s_z k_z}$.
The effect propagates through to the symmetrized function
$u^\prime(\bk)$,
which is therefore non-zero
only for the $27$ wavevectors $\bk$ all of whose components are multiples of $4$.
The symmetrized function $u^\prime(\bk)$
can be computed by averaging the values of $u^\prime_i(\bk)$
at the $27$ values of $\bk$
into the $4$ distinct bins with
$0 \leq k_x \leq k_y \leq k_z \leq 4$
and
each component a multiple of $4$.

One more step,
adjoining translations by an eighth of a box,
reduces the problem of finding the minimum variance solution
to a triviality.
After adjoining translations by an eighth of a box,
the symmetrized function $u^\prime(\bk)$
vanishes except at $\bk = \zero$.
The function to be minimized,
the right hand side of equation~(\ref{DXhat2u}),
is therefore identically equal to $1$,
and any arbitrary weighting therefore yields a minimum variance solution.
Though amusing,
the result is not terribly useful,
because it involves a vast number, $48 \times 8^3 = 24{,}576$,
of weightings.
Physically,
if there are enough weightings,
then together they exhaust the information
about the covariance of power,
however badly crafted the weightings may be.
As will be seen in \S\ref{theweightings},
there are much simpler solutions that achieve the absolute
minimum possible variance, for which
the right hand side of equation~(\ref{DXhat2u})
equals $1$,
with far fewer weightings.

The argument above
has shown that the problem of finding the minimum variance solution
for $w_i(\bk)$
attains its simplest non-trivial form
if the weightings are generated from a representative weighting
by rotations, reflections, and translations by quarter of a box,
a total of
$48 \times 4^3 = 3{,}072$
symmetries.
In this case,
the weighting-dependent factor in
the variance of covariance of power,
the right hand side of equation~(\ref{DXhat2u}),
becomes a rational function,
a ratio of two $8$th order polynomials
in the $27$ Fourier coefficients $w_i(\bk)$,
the numerator being a sum
$\sum u^\prime(\bk)^2$ of squares of $4$ quartics,
and the denominator $u^\prime(\zero)^2$ the square of a quartic.
It is this function that we minimize in \S\ref{theweightings}
to find a best set of weightings.

The minimum variance solution is
independent of the overall normalization of the coefficients $w_i(\bk)$,
since the quantity being minimized,
the ratio on the right hand side of equation~(\ref{DXhat2u}),
is independent of the normalization of $w_i(\bk)$.
Once the minimum variance solution for the coefficients $w_i(\bk)$
has been found,
the coefficients can be renormalized to satisfy the
normalization condition~(\ref{wnorm})
that ensures that the estimates $\phat_i(k)$
of the shell-averaged power spectra of weighted densities
are estimates of the true shell-averaged power $p(k)$,
equations~(\ref{pi}) and (\ref{piapprox}).

\subsection{Minimum variance weightings}
\label{theweightings}

The previous subsection, \S\ref{minvar},
described how to obtain the coefficients $w_i(\bk)$
that minimize the expected variance
of the estimate of covariance of shell-averaged power
that comes from averaging over an ensemble of weightings
that contain only combinations of fundamental modes,
and that are symmetrically related to each other by
rotations, reflections, and translations by quarter of a box.

\wfig

Numerically,
we find not one but three separate sets of minimum variance weightings
(with hindsight, the sets are simple enough that they might
perhaps have been found without resort to numerics).
Each set consists of symmetrical transformations of
a weighting generated by a single mode,
namely
$\{1,0,0\}$, $\{1,1,0\}$, and $\{1,1,1\}$
respectively for each of the three sets.
Because each individual weighting has a rather high degree of symmetry,
each set has far fewer than the
$48 \times 4^3 = 3{,}072$ weightings
expected if all symmetrical transformations yielded distinct weightings.
Each of the three sets is generated by the weighting
\begin{equation}
\label{wkminvar}
  w_i(\bk)
  =
  \left\{
  \begin{array}{ll}
    \e^{\pm \im \upi / 8} / \sqrt{2} & \mbox{if } \bk = \pm \bk_i \\
    0 & \mbox{otherwise}
  \end{array}
  \right.
\end{equation}
where $\bk_i$ is one of the three possibilities
\begin{equation}
\label{kminvar}
  \bk_i
  =
  \left\{
  \begin{array}{ll}
    \{1,0,0\} & \mbox{set one: $12$ weightings} \\
    \{1,1,0\} & \mbox{set two: $24$ weightings} \\
    \{1,1,1\} & \mbox{set three: $16$ weightings.}
  \end{array}
  \right.
\end{equation}
In real space, the weighting $w_i(\br)$
corresponding to $w_i(\bk)$ of equation~(\ref{wkminvar}) is
\begin{equation}
\label{wrminvar}
  w_i(\br)
  =
  \sqrt{2} \cos \Bigl[ 2\upi \Bigl( \bk_i . \br + \frac{1}{16} \Bigr) \Bigr]
  \ .
\end{equation}
The complete set of
$12$ ($24$, $16$)
weightings for each set is obtained
as follows.
In set one (two, three),
a factor of $6$ ($12$, $8$)
comes from the cubic (dodecahedral, octohedral)
symmetry of permuting and reflecting the components
$k_x$, $k_y$, $k_z$ of $\bk$,
or equivalently the components
$x$, $y$, $z$ of $\br$.
A further factor of $2$ comes from
multiplying $w_i(\pm\bk)$ by $\pm\im$,
equivalent to translating by quarter of a box,
or $1/16 \rightarrow 5/16$
in equation~(\ref{wrminvar}).

The three minimum variance solutions are absolute minimum variance,
in the sense that each set not only minimizes
the expression on the right hand side of equation~(\ref{DXhat2u}),
but it solves
$u^\prime(\bk) = 0$ for $\bk \neq \zero$.
This means that
it is impossible to find better solutions
in which all the weightings are symmetrically related to each other,
which is the condition
under which equation~(\ref{DXhat2u}) was derived.

With the minimum variance solutions in hand,
it is possible to go back and examine
the covariance
$\langle \Delta\phat^\prime_i \Delta\phat^\prime_j \rangle$,
equation~(\ref{DphatiDphatjfp}),
between estimates of power from different weightings $i$ and $j$,
either within the same set, or across two different sets.
Estimates of power between two different sets are uncorrelated:
the covariance $\langle \Delta\phat^\prime_i \Delta\phat^\prime_j \rangle$
is zero if $i$ and $j$ are drawn from two different sets.
If on the other hand the weightings $i$ and $j$ are drawn from the same set,
then it turns out that only half of the weightings,
the $6$ ($12$, $8$) weightings related by the
cubic (dodecahedral, octohedral) symmetry of
permuting and reflecting $k_x$, $k_y$, $k_z$,
yield distinct estimates of deviation in power.
The covariance matrix
$\langle \Delta\phat^\prime_i \Delta\phat^\prime_j \rangle$
of estimates of power
between the $6$ ($12$, $8$)
cubically (dodecahedrally, octohedrally) related weightings
is proportional to the unit matrix.
However, translating a weighting by quarter of a box,
$w_j(\pm\bk) = \pm\im \, w_i(\pm\bk)$,
yields an estimate of deviation of power
that is minus that of the original weighting,
$\Delta\phat^\prime_j = - \Delta\phat^\prime_i$.
Actually, this is exactly true
only if the slowly-varying and thick-shell approximations
are exactly true
(of course,
the thick-shell approximation is never exactly true).
Thus
translating a weighting by quarter of a box
should yield an estimate of deviation in power
that is highly anti-correlated with the original;
which should provide a useful check of the procedure.

Translating a weighting by half a box simply changes its sign,
$w_j(\pm\bk) = - w_i(\pm\bk)$.
This yields an estimate of deviation of power
that equals exactly (irrespective of approximations)
that of the original weighting,
so yields no distinct estimate of deviation in power.
These redundant translations by half a box
have already been omitted from the set of $12$ ($24$, $16$) weightings.

The value of $f^\prime$,
the factor that converts, equation~(\ref{Xhati}),
estimates $\Xhat_i$ of the covariance of power from a weighted density field
to an
estimate of the true covariance of power is
\begin{equation}
\label{fpminvar}
  f^\prime = 1/2
\end{equation}
the same factor for each of the three sets.

The expected covariance matrix
$\langle \Delta\Xhat_i \Delta\Xhat_j \rangle$
of estimates of covariance of power
equals $g^\prime_{ij}$ times the true covariance of covariance of power,
according to equation~(\ref{Dphati2Dphatj2g}).
The factors $g^\prime_{ij}$,
equation~(\ref{gpiju}),
are
\begin{equation}
\label{gpijminvar}
  g^\prime_{ij}
  =
  \left\{
  \begin{array}{ll}
    3/8 & \mbox{if $\bk_i = \bk_j$} \\
    1/8 & \mbox{if $\bk_i = - \bk_j$} \\
    1/4 & \mbox{otherwise.}
  \end{array}
  \right.
\end{equation}
Equation~(\ref{gpijminvar})
is valid for weightings $i$, $j$
both within the same set and across different sets.
The case $\bk_i = \bk_j$ in equation~(\ref{gpijminvar})
occurs not only when $i = j$,
but also when the weightings $i$ and $j$
are related by translation by quarter of a box.
The case $\bk_i = - \bk_j$ in equation~(\ref{gpijminvar})
occurs not only when the weightings $i$ and $j$
are parity conjugates of each other,
but also when they are parity confugates
translated by quarter of a box.

The factors $g^{\prime\prime}_{ij}$,
equation~(\ref{gppiju}),
which relate the covariance
$\langle (\Xhat_i{-}\Xhat) (\Xhat_j{-}\Xhat) \rangle$
of estimates $\Xhat_i$
relative to their measured mean $\Xhat$, equation~(\ref{Xhat}),
as opposed to their expected mean $X$, equation~(\ref{Xi}),
are
\begin{equation}
\label{gppijminvar}
  g^{\prime\prime}_{ij}
  =
  \left\{
  \begin{array}{ll}
    1/8 & \mbox{if $\bk_i = \bk_j$} \\
    - 1/8 & \mbox{if $\bk_i = - \bk_j$} \\
    0 & \mbox{otherwise.}
  \end{array}
  \right.
\end{equation}

An estimate of the uncertainty in the estimate $\Xhat$
can be deduced by measuring the variance
$N^{-1} \sum_i (\Xhat_i{-}\Xhat)^2$
in the fluctuations
about the measured mean $\Xhat$.
There is of course no point in attempting to estimate the uncertainty from
$N^{-2} \sum_{ij} \Delta\Xhat_i \Delta\Xhat_j$,
which is identically zero.
The true variance
$\langle \Delta\Xhat^2 \rangle$
can be estimated
from the measured variance
$N^{-1} \sum_i (\Xhat_i{-}\Xhat)^2$
by
\begin{equation}
\label{DXhat2est}
  \langle \Delta\Xhat^2 \rangle
  \approx
  {2 \over N}
  \sum_i (\Xhat_i - \Xhat)^2
\end{equation}
in which the factor of $2$ comes from
(but note the caveat at the end of \S\ref{notquiteweightings})
\begin{equation}
\label{ggminvar}
  {N^{-2} \sum_{ij} g^\prime_{ij} \over N^{-1} \sum_{i} g^{\prime\prime}_{ii}}
  =
  {1/4 \over 1/8}
  =
  2
\end{equation}
which corrects for the neglected covariance in the measured variance.

\subsection{Minimum variance weightings for strategy two}
\label{theweightings2}

The minimum variance weightings derived above assumed,
for definiteness,
strategy one,
in which the deviation
$\Delta\phat^\prime_i(\bk)$
in power is taken
to be relative to the power spectrum of the unweighted density,
equation~(\ref{Dphati1}),
An alternative strategy, strategy two,
is to take the deviation
$\Delta\phat^\prime_i(\bk)$
in power
to be relative to the average of the power spectra of the weighted densities,
equation~(\ref{Dphati2}).
Strategy two yields an estimate of covariance of power that
has potentially less systematic bias,
but potentially greater statistical uncertainty.

As it happens,
the minimum variance solution for strategy one,
\S\ref{theweightings},
proves also to solve the minimum variance problem for strategy two.
Thus the minimum variance solution weightings are the same for both strategies.
Mathematically,
expectation values of covariances for the two methods
differ in that $v^\prime_i(\bk)$
is given for strategy one by equation~(\ref{vpi1}),
and for strategy two by equation~(\ref{vpi2}).
However,
for the minimum variance weightings $w_i(\bk)$ of strategy one,
equation~(\ref{wkminvar}) and its symmetrical transformations,
it turns out that
$N^{-1} \sum_i v_i(\bk)$,
the term subtracted from $v_i(\bk)$ in strategy two,
equation~(\ref{vpi2}),
is equal to $v_i(\zero)$ if $\bk = \zero$,
and zero otherwise.
This is exactly the same as
the term subtracted from $v_i(\bk)$ in strategy one,
equation~(\ref{vpi1}).
It follows that
$v^\prime_i(\bk)$ is the same for the two strategies.

Although the minimum variance set of weightings is the
same for both strategies,
the two strategies will in general yield
different estimates of the covariance of power.

\subsection{More minimum variance weightings}
\label{minvarmore}

The three minimum variance sets of weightings found (numerically)
in \S\ref{theweightings}
all take the same form, equation~(\ref{wkminvar}),
differing only in that they are generated by a different single mode,
with wavevectors
$\{1,0,0\}$, $\{1,1,0\}$, and $\{1,1,1\}$
respectively.
One can check that
the result generalizes to higher order weightings,
in which the wavevector $\bk_i$ in equation~(\ref{wkminvar})
is any wavevector with integral components
(such as $\{2,0,0\}$, $\{2,1,0\}$, and so on).
That is, for any wavevector
$\bk_i$ with integral components,
the weightings generated
from the weighting of equation~(\ref{wkminvar})
by rotations, relections, and translations by quarter of a box,
form a minimum variance set.
All the results of \S\ref{theweightings} (and \S\ref{theweightings2})
carry through essentially unchanged.
In particular, all equations~(\ref{wrminvar})--(\ref{ggminvar})
remain the same.

The disadvantage of including higher order weightings
is that the estimates $\Xhat_i$
of the covariance of power
become increasingly inaccurate as the wavenumber
$|\bk_i|$ of the weighting increases,
because
the slowly-varying approximations~(\ref{Papprox}) and (\ref{Tapprox}),
and the broad-shell approximation~(\ref{broadshellapprox}),
become increasingly poor
as $|\bk_i|$ increases.

The advantage of including higher order weightings is that
the more weightings, the better the statistical estimate,
at least in principle.
However,
the gain from more weightings is not as great as one might hope.
The Cram\'{e}r-Rao inequality
(\citealt{KS67};
see e.g.\ \citealt{H05} for a pedagogical derivation)
states that the inverse variance of the best possible unbiassed estimate
$\Xhat$ of the parameter $X$
must be less than or equal to the Fisher information $F$
(see \citealt{TTH97})
in the parameter $X$
\begin{equation}
\label{CramerRao}
  \langle \Delta\Xhat^2 \rangle^{-1}
  \leq
  F
  \equiv
  - \Bigl\langle {\partial^2 \ln {\cal L} \over \partial X^2} \Bigr\rangle
\end{equation}
where ${\cal L}$ is the likelihood function.
To the extent that
the estimates $\Xhat_i$ are Gaussianly distributed
(that is,
the likelihood function is a Gaussian in the estimates $\Xhat_i$
\begin{equation}
  {\cal L}
  \simpropto
  \exp \Bigl[ - \frac{1}{2}
    \sum_{ij} \langle \Delta\Xhat_i \Delta\Xhat_j \rangle^{-1}
    (\Xhat_i - X ) ( \Xhat_j - X)
  \Bigr]
\end{equation}
with covariance
$\langle \Delta\Xhat_i \Delta\Xhat_j \rangle$
independent of $X$),
which could be a rather poor approximation,
the Fisher information $F$ in the parameter $X$
approximates the sum of the elements of the inverse covariance matrix,
\begin{equation}
\label{Fisher}
  F
  \approx
  \sum_{ij} \langle \Delta\Xhat_i \Delta\Xhat_j \rangle^{-1}
 \ .
\end{equation}
In the present case,
the covariance matrix
$\langle \Delta\Xhat_i \Delta\Xhat_j \rangle$
is proportional to $g^\prime_{ij}$,
so in approximation that $\Xhat_i$ are Gaussianly distributed,
the Fisher information $F$ is proportional to
\begin{equation}
\label{Fisherg}
  F
  \simpropto
  \sum_{ij} {g^\prime_{ij}}^{-1}
  \ .
\end{equation}
With the coefficients $g^\prime_{ij}$
given by equation~(\ref{gpijminvar}),
the quantity on the right hand side of equation~(\ref{Fisherg})
proves to be a constant,
independent of the number of estimates $\Xhat_i$
\begin{equation}
\label{Fisherminvar}
  \sum_{ij} {g^\prime_{ij}}^{-1}
  =
  4
  \ .
\end{equation}
This constancy of the Fisher information $F$
with respect to the number of estimates
suggests that
there is no gain at all in adjoining more and more estimates.
However, this conclusion
is true only to the extent, firstly, that the slowly-varying
and broad-shell approximations are good,
and, secondly, that
the estimates $\Xhat_i$ are Gaussianly distributed,
neither of which assumptions necessarily holds.
All one can really conclude
is that the gain in statistical accuracy
from including more estimates is likely to be limited.

There is however another important consideration
besides the accuracy of the estimate of the covariance matrix of power:
it is desirable
that the estimated covariance matrix be,
like the true covariance matrix,
strictly positive definite,
that is, it should have no zero (or negative) eigenvalues.
As noted by
\cite{PS05},
if a matrix is estimated as an average over $N$ estimates,
then its rank can be no greater than $N$.
Thus,
to obtain a positive definite
covariance matrix of power for $N$ shells of wavevector,
at least $N$ distinct estimates $\Xhat_i$
are required.

In \S\ref{recommend} below
we recommend estimating the covariance of power from
an ensemble of $12 + 24 + 16 = 52$ weightings.
This will yield a positive definite covariance matrix
only if the covariance of power is estimated
over no more than $52$ shells of wavenumber.
Since, as noted in \S\ref{theweightings},
weightings related by translation by quarter of a box
yield highly anti-correlated estimates of power,
hence highly correlated estimates of covariance of power,
a more conservative approach would be to consider
that the $52$ weightings yield
only $26$ effectively distinct estimates of covariance of power,
so that the covariance of power can be estimated
over no more than $26$ shells of wavenumber.
If
(strategy two)
the deviation of power
is measured relative to
the measured mean over symmetrically related weightings,
a (slightly) different mean for each of the $3$
sets of weightings,
then $3$ degrees of freedom are lost,
and the covariance of power can be estimated
over no more than $52 - 3 = 49$ shells of wavenumber,
or more conservatively
over no more than $26 - 3 = 23$ shells of power.

\subsection{Recommended strategy}
\label{recommend}

Here is a step-by-step recipe for applying the weightings method
to estimate the covariance of power from a periodic simulation.
\begin{enumerate}
\item
Select the weightings $w_i$.
We recommend the minimum variance sets of weightings
given by equation~(\ref{wkminvar})
and its symmetrical transformations.
If the weightings are restricted to contain
only combinations of fundamental modes,
then there are three such sets of weightings,
equation~(\ref{kminvar}),
and the three sets together provide
$N = 12 + 24 + 16 = 52$ distinct weightings.
\item
For each weighting,
measure the shell-averaged power spectrum $\phat_i(k)$
of the weighted density field,
equations~(\ref{phati}) and (\ref{Phati}).
\item
For each weighting,
evaluate the deviation $\Delta\phat_i(k)$
in the shell-averaged power
as the difference between $\phat_i(k)$
and,
either (strategy one)
the shell-averaged power $\phat(k)$ of the unweighted density,
or (strategy two)
the mean
$N^{-1} \sum_i \phat_i(k)$
over symmetrically related weightings.
The advantage of strategy one is that
the statistical error is potentially smaller,
whereas the advantage of strategy two is that
the systematic bias is potentially smaller.
In strategy two,
it makes sense to subtract the mean separately
for each symmetrically related set of weightings,
because the systematic bias is (slightly) different
for each set.
We recommend trying both strategies one and two,
and checking that they yield consistent results.
\item
Estimate the covariance matrix of shell-averaged power
from the average over all $N$ ($52$) weightings
\begin{equation}
\label{Xest}
  \bigl\langle \Delta\phat(k_1) \Delta\phat(k_2) \bigr\rangle_{\rm est}
  =
  {2 \over N}
  \sum_{i} \Delta\phat_i(k_1) \Delta\phat_i(k_2)
  \ .
\end{equation}
The factor of $2$ in equation~(\ref{Xest})
is $1/f^\prime = 2$, equation~(\ref{fpminvar}),
necessary to convert the average over weightings
to an estimate of the true covariance of power,
equation~(\ref{Xhati}).
\end{enumerate}

\vartwofig

\section{Beat-coupling}
\label{beatcoupling}

This paper should have ended at this point.
Unfortunately,
numerical tests,
described in detail in the companion paper
\citep{RH06}
revealed a serious problem.



Figure~\ref{vartwofig}
shows the problem.
It shows the median and quartiles
of variance of power measured by the weightings method
in each of 25 ART $\Lambda$CDM simulations of $128 \, h^{-1} \Mpc$ box size,
compared to the variance of power measured over the ensemble of
the same 25 simulations.
Although the two methods agree at linear scales,
the weightings method gives a systematically larger variance at nonlinear scales.
The discrepancy reaches almost an order of magnitude
at the smallest scales measured,
$k \sim 5 \, h^{-1} \Mpc$.
The reader is referred to \citet{RH06} for details
of the simulations and their results.

This section diagnoses and addresses the problem.
The next section, \S\ref{discussion},
discusses the problem
and its relevance to observations.

\subsection{The cause of the problem: beat-coupling}
\label{cause}

The physical cause of the problem
illustrated in Figure~\ref{vartwofig}
traces to a nonlinear coupling of products of Fourier modes
closely spaced in wavenumber
to the large-scale beat mode between them.
This beat-coupling, as we refer to it,
occurs only when power is measured
from Fourier modes with a finite spread in wavevector,
and therefore appears in the weightings method
(and in observations -- see \S\ref{observation} below)
but not in the ensemble method.
The beat-coupling is surprisingly large,
to the point that,
as seen in Figure~\ref{vartwofig},
it actually dominates the variance of power at nonlinear scales.

More specifically,
in the ensemble method,
the power spectrum of a periodic simulation
is measured from the variance
$\Delta\rho(\bk) \Delta\rho(-\bk)$
of Fourier modes.
In the weightings method
on the other hand,
the power spectrum
receives contributions not only from the variance,
but also from the covariance
$\Delta\rho(\bk) \Delta\rho(-\bk{-}\bvarepsilon)$
between modes a small wavevector $\bvarepsilon$ apart.
This covariance vanishes in the mean,
but it couples to large-scale modes
$\Delta\rho(\bvarepsilon)$
through quadratic nonlinearities.
That is,
the correlation between the product
$\Delta\rho(\bk) \Delta\rho(-\bk{-}\bvarepsilon)$
and the large-scale mode
$\Delta\rho(\bvarepsilon)$
is the bispectrum
\begin{equation}
  \langle \Delta\rho(\bk) \Delta\rho(-\bk{-}\bvarepsilon) \Delta\rho(\bvarepsilon) \rangle
  =
  B(\bk, -\bk{-}\bvarepsilon, \bvarepsilon)
  \ .
\end{equation}
The bispectrum is zero for Gaussian fluctuations,
but is driven away from zero by nonlinear gravitational growth.

\subsection{Tetrahedron}
\label{tetrahedron}

The place where, prior to this section, we inadvertently
discarded the large-scale beat-coupling, is equation~(\ref{Tapprox}),
where we made the seemingly innocent approximation
that the trispectrum
$T(\bk_1, \bk_2, \bk_3, \bk_4)$
is a slowly varying function of what appears to be its arguments,
$\bk_1$ to $\bk_4$.
This assumption is false,
as we now show.

For a statistically isotropic field (as considered in this paper),
the trispectrum depends on six scalar arguments.
This follows from the fact that
a spatial configuration of four points is determined by
the six lengths of
the sides of the tetrahedron whose vertices are the four points.
In Fourier space,
the configuration
is an object four of whose sides are
equal to the wavevectors $\bk_1$ to $\bk_4$.
The object forms a closed tetrahedron
(because $\sum_i \bk_i = \zero$),
whose shape
is determined by the six lengths of the sides of the tetrahedron.

\tetrahedronfig

Figure~\ref{tetrahedronfig}
illustrates the configuration of interest in the present paper,
that for the trispectrum in equation~(\ref{DphatiDphatj}).
Rewritten as a function of six scalar arguments,
the trispectrum
of equation~(\ref{DphatiDphatj})
is
\begin{eqnarray}
\label{Tsix}
\lefteqn{
  T(\bk_1 {-} \bk_1^\prime, -\bk_1 {-} \bk_1^{\prime\prime}, \bk_2 {-} \bk_2^\prime, -\bk_2 {-} \bk_2^{\prime\prime})
  =
}
&&
\\
\nonumber
\lefteqn{
  T\bigl( | \bk_1 {-} \bk_1^\prime | , | \bk_1 {+} \bk_1^{\prime\prime} | ,  | \bk_2 {-} \bk_2^\prime | , | \bk_2 {+} \bk_2^{\prime\prime} | ,
  | \bk_1 {-} \bk_2 {-} \bk_1^\prime {-} \bk_2^{\prime\prime} | , \varepsilon \bigr)
}
&&
\end{eqnarray}
where the wavevector $\bvarepsilon$ is defined by
\begin{equation}
\label{epsilon}
  \bvarepsilon
  \equiv
  - ( \bk^\prime_1 + \bk^{\prime\prime}_1 )
  =
  \bk^\prime_2 + \bk^{\prime\prime}_2
\end{equation}
which is small but not necessarily zero.
The invalid approximation~(\ref{Tapprox}) is equivalent to approximating
\begin{eqnarray}
\label{Tapprox1}
\lefteqn{
  T\bigl( | \bk_1 {-} \bk_1^\prime | , | \bk_1 {+} \bk_1^{\prime\prime} | ,  | \bk_2 {-} \bk_2^\prime | , | \bk_2 {+} \bk_2^{\prime\prime} | ,
  | \bk_1 {-} \bk_2 {-} \bk_1^\prime {-} \bk_2^{\prime\prime} | , \varepsilon \bigr)
}
\nonumber
\\
&\approx&
  T( k_1 , k_1 , k_2 , k_2 , | \bk_1 {-} \bk_2 | , 0 )
  \ .
\end{eqnarray}
The problem with this approximation is apparent.
Although primed wavenumbers are small compared to unprimed ones,
so that the approximation in the first five arguments is reasonable,
in the last argument it is not valid
to approximate a finite wavenumber $\varepsilon$,
however small, by zero.
A valid approximation is, rather,
\begin{eqnarray}
\label{Tapprox2}
\lefteqn{
  T\bigl( | \bk_1 {-} \bk_1^\prime | , | \bk_1 {+} \bk_1^{\prime\prime} | ,  | \bk_2 {-} \bk_2^\prime | , | \bk_2 {+} \bk_2^{\prime\prime} | ,
  | \bk_1 {-} \bk_2 {-} \bk_1^\prime {-} \bk_2^{\prime\prime} | , \varepsilon \bigr)
}
\nonumber
\\
&\approx&
  T( k_1 , k_1 , k_2 , k_2 , | \bk_1 {-} \bk_2 | , \varepsilon )
  \ .
\end{eqnarray}

As an example of the large-scale beat-coupling contributions
to the trispectrum
that arise from the beat wavevector $\bvarepsilon$,
consider perturbation theory.

\subsection{Perturbation theory}
\label{PT}

In perturbation theory (PT),
the  trispectrum can be split into snake and star
contributions
(\citealt{SZH99}; \citealt{SS05})
\begin{eqnarray}
\label{Tab}
\lefteqn{
  T(\bk_1, \bk_2, \bk_3, \bk_4)
  =
}
&&\!\!\!\!\!\!
\nonumber
\\
&&\!\!\!\!\!\!
  4 \, P(k_1) P(k_2)
  P(k_{13}) F_2(\bk_1,-\bk_{13}) F_2(\bk_2,\bk_{13})
\nonumber
\\
&&\!\!\!\!\!\!
  \ 
  \mbox{}
  +
  \mbox{cyclic (12 snake terms)}
\nonumber
\\
&&\!\!\!\!\!\!
  \mbox{}
  +
  P(k_1) P(k_2) P(k_3)
  \bigl[
  F_3(\bk_1,\bk_2,\bk_3) + \mbox{perm.~(6 terms)}
  \bigr]
\nonumber
\\
&&\!\!\!\!\!\!
  \ 
  \mbox{}
  +
  \mbox{cyclic (4 star terms)}
\end{eqnarray}
where
$\bk_{ij} \equiv \bk_i + \bk_j$,
and the second-order PT kernel $F_2$ is given by
\begin{equation}
  F_2(\bk_1,\bk_{2})
  =
  \frac{5}{7}
  + \frac{x}{2} \left( \frac{k_1}{k_2} + \frac{k_2}{k_1} \right)
  + \frac{2}{7}\ x^2
\label{F2}
\end{equation}
with $x \equiv \hat{\bk}_1 \cdot \hat{\bk}_2$.

In the case of interest,
where the trispectrum is that of equation~(\ref{Tsix}),
4 of the 12 snake terms produce a coupling to large scales,
those where the beat wavenumber
$k_{13}$
in equation~(\ref{Tab})
is small.
In the (valid) approximation~(\ref{Tapprox2}),
the pertinent PT trispectrum is
\begin{eqnarray}
\label{TapproxPT}
\lefteqn{
  T\bigl( | \bk_1 {-} \bk_1^\prime | , | \bk_1 {+} \bk_1^{\prime\prime} | ,  | \bk_2 {-} \bk_2^\prime | , | \bk_2 {+} \bk_2^{\prime\prime} | ,
  | \bk_1 {-} \bk_2 {-} \bk_1^\prime {-} \bk_2^{\prime\prime} | , \varepsilon \bigr)
}
\nonumber
\\
&\approx&
  T( k_1 , k_1 , k_2 , k_2 , | \bk_1 {-} \bk_2 | , 0 )
\nonumber
\\
&&
  \mbox{}
  +
  16 \, P(k_1) P(k_2)
  P(\varepsilon) F_2(\bk_1,-\bvarepsilon) F_2(\bk_2,\bvarepsilon)
\end{eqnarray}
in which the term on the last line represents the large-scale beat-coupling contribution
incorrectly ignored by the approximation~(\ref{Tapprox1}).
In equation~(\ref{DphatiDphatj})
for the covariance of shell-averaged power,
this trispectrum, equation~(\ref{TapproxPT}),
is angle-averaged over the directions
of $\bk_1$ and $\bk_2$.
The angle-averaged second-order PT kernel is
\begin{equation}
  \int F_2(\bk,\bvarepsilon) \, {do_\bk \over 4\upi}
  =
  \frac{17}{21}
\end{equation}
and it follows that the last line of equation~(\ref{TapproxPT}),
when angle-averaged, is
$16 (17/21)^2 P(k_1) P(k_2) P(\varepsilon)$.

Following the same arguments that led from equation~(\ref{DphatiDphatj})
to equation~(\ref{DphatiDphatjf}),
and then to equation~(\ref{DphatiDphatjfp}),
but with the beat-coupling term now correctly retained in the trispectrum,
one finds that 
equation~(\ref{DphatiDphatjfp})
for the expected covariance of shell-averaged power
spectra of weighted densities
is modified to
\begin{eqnarray}
\label{DphatiDphatjfLPT}
\lefteqn{
  \left\langle \Delta\phat^\prime_i(k_1) \Delta\phat^\prime_j(k_2) \right\rangle
  \approx
  f^\prime_{ij}
  \left\langle \Delta\phat(k_1) \Delta\phat(k_2) \right\rangle
}
&&
\nonumber
\\
&&
  \qquad
  \mbox{}
  +
  4 \, R_a P(k_1) P(k_2)
  \sum_\bk v^\prime_i(\bk) v^\prime_j(-\bk) P(k)
\end{eqnarray}
where
$v^\prime_i(\bk)$ is defined by equations~(\ref{vi})
and (\ref{vpi1}) or (\ref{vpi2}),
and the constant $R_a$ is
\begin{equation}
\label{RaPT}
  R_a = 4 \left( \frac{17}{21} \right)^2 \approx 2.62
  \ .
\end{equation}
The reason for writing equation~(\ref{DphatiDphatjfLPT})
in this form, with the constant $R_a$ separated out,
is that,
as will be seen in \S\ref{hierarchical},
the same expression remains valid in the hierarchical model,
but with $R_a$ the 4-point hierarchical snake amplitude.

Figure~\ref{vartwofig}
includes lines showing the predicted PT result
for the variance of shell-averaged power of weighted density,
equation~(\ref{DphatiDphatj}),
both with (solid lines)
and without (dashed lines)
beat-coupling.
The PT variance with beat-coupling was obtained
by numerically integrating the PT expression~(\ref{Tab})
for the trispectrum~(\ref{Tsix}) in
equation~(\ref{DphatiDphatj})
(that is, without making the approximations~(\ref{Tapprox2})
or (\ref{DphatiDphatjfLPT})),
with the minimum variance weightings~(\ref{wkminvar}),
and then multiplying by the factor $1/f^\prime = 2$,
equation~(\ref{fpminvar}).
From this
the PT variance without beat-coupling was obtained
by setting $P(\varepsilon) = 0$.
The variance without beat-coupling agreed well with a direct
PT evaluation of equation~(\ref{DphatDphat}).

Figure~\ref{vartwofig}
shows that
the beat-coupling contribution
predicted by perturbation theory
seems to account reasonably well for
the extra variance that appears at nonlinear scales
in the weightings versus the ensemble method.

We will return to equation~(\ref{DphatiDphatjfLPT})
in \S\ref{largescale} below,
but first consider the hierarchical model
as a prototype of the trispectrum beyond perturbation theory.

\subsection{Hierarchical model}
\label{hierarchical}

Perturbation theory is valid only in the translinear regime.
The behaviour of the trispectrum in the fully nonlinear regime
is less well understood.
Available observational and $N$-body evidence
(\citealt{CBH96}; \citealt{HG99}; \citealt{SF99}; \citealt{Baugh04}; \citealt{Croton04})
is consistent
with a hierarchical model of higher order correlations.
In the hierarchical model
\citep{Peebles80},
the trispectrum is a sum of snake and star terms
\begin{eqnarray}
\label{Thier}
\lefteqn{
  T(\bk_1, \bk_2, \bk_3, \bk_4) =
}
&&\!\!\!\!\!\!
\nonumber
\\
&&\!\!\!\!\!\!
  R_a \bigl[ P(k_1) P(k_2) P(k_{13}) + \mbox{cyclic~(12 snake terms)} \bigr]
\nonumber
\\
&&\!\!\!\!\!\!
  \mbox{}
  +
  R_b \bigl[ P(k_1) P(k_2) P(k_3) + \mbox{cyclic~(4 star terms)} \bigr]
  \ .
\end{eqnarray}
The PT trispectrum,
equation~(\ref{Tab}),
shows a hierarchical structure
with hierarchical amplitudes $R_a$ and $R_b$
that are not constant,
but rather depend on the shape of the trispectrum tetrahedron.
At highly nonlinear scales,
Scoccimarro \& Frieman (1999) suggested an ansatz,
dubbed hyperextended perturbation theory (HEPT),
that the hierarchical amplitudes
go over to the values predicted by perturbation theory
for configurations collinear in Fourier space.
For power law power spectra $P(k) \propto k^n$,
HEPT predicts 4-point amplitudes
\begin{equation}
\label{Q4}
  R_a = R_b = {54 - 27\ 2^n + 2\ 3^n + 6^n \over 2 \, (1 + 6\ 2^n + 3\ 3^n + 6\ 6^n)}
 \ .
\end{equation}
As pointed out by \citet{SF99}
and
\cite{H00},
HEPT is not entirely consistent because it predicts
a covariance of power
$\langle \Delta\phat(k_1) \Delta\phat(k_2) \rangle$
that violates the Schwarz inequality when $k_1 \gg k_2$.

In the hierarchical model with constant hierarchical amplitudes,
4 of the 12 snake terms produce a coupling to large
scales in the trispectrum of interest,
equation~(\ref{Tsix}).
In the (valid) approximation~(\ref{Tapprox2}),
the hierarchical trispectrum is
\begin{eqnarray}
\label{Tapproxhier}
\lefteqn{
  T\bigl( | \bk_1 {-} \bk_1^\prime | , | \bk_1 {+} \bk_1^{\prime\prime} | ,  | \bk_2 {-} \bk_2^\prime | , | \bk_2 {+} \bk_2^{\prime\prime} | ,
  | \bk_1 {-} \bk_2 {-} \bk_1^\prime {-} \bk_2^{\prime\prime} | , \varepsilon \bigr)
}
\nonumber
\\
&\approx&
  T( k_1 , k_1 , k_2 , k_2 , | \bk_1 {-} \bk_2 | , 0 )
\nonumber
\\
&&
  \mbox{}
  +
  4 \, R_a \, P(k_1) P(k_2)
  P(\varepsilon)
\end{eqnarray}
in which the term on the last line represents the large-scale beat-coupling contribution.

The hierarchical trispectrum~(\ref{Tapproxhier})
looks similar to (slightly simpler than)
the PT trispectrum~(\ref{TapproxPT}).
Following the same arguments as before,
one recovers the same expression~(\ref{DphatiDphatjfLPT})
for the expected covariance of shell-averaged power
spectra of weighted densities.

\subsection{Covariance of shell-averaged power spectra including large-scale coupling}
\label{largescale}

Suppose that either perturbation theory, \S\ref{PT},
or the hierarchical model, \S\ref{hierarchical},
offers a reliable guide to
the coupling of the nonlinear trispectrum to large scales,
so that equation~(\ref{DphatiDphatjfLPT}) is a good approximation
to the expected covariance of shell-averaged power spectra
of weighted densities.

Make the further assumption that the power spectrum
is approximately constant
over the large-scale wavevectors
represented in $v^\prime_i(\bk)$
\begin{equation}
\label{PapproxL}
  P(k) \approx P(2 k_b) = \mbox{constant}
  \quad
  \mbox{for $v_i^\prime(\bk) \neq 0$}
\end{equation}
where $k_b$ is the wavenumber at the box scale.
The factor $2$ in $2 k_b$ in equation~(\ref{PapproxL})
appears as a reminder that
the wavevectors $\bk$ in $v^\prime_i(\bk)$ are,
equations~(\ref{vi}) and (\ref{vpi1}) or (\ref{vpi2}),
sums of pairs of wavenumbers $\bk^\prime$
represented in the weighting $w_i(\bk^\prime)$.
For example, if the weightings are taken
to be the minimum variance weightings
given by equation~(\ref{wkminvar}),
then $k_b = k_i$
where $k_i$ is the wavenumber of the weighting.
Approximation~(\ref{PapproxL})
is in the same spirit as, but distinct from,
the earlier approximation~(\ref{Papprox})
that the power spectrum is a slowly varying function.
Note that equation~(\ref{PapproxL}) does {\em not\/}
require that $P(0) \approx P(2 k_b)$
(which would certainly not be correct,
because $P(0) = 0$),
because $v^\prime_i(\zero)$ is zero,
which is true a priori in strategy one,
equation~(\ref{vpi1}),
and ends up being true a posteriori in strategy two,
equation~(\ref{vpi2}),
by the argument in \S\ref{theweightings2}.

In the approximation~(\ref{PapproxL}),
the summed expression on the right hand side of
equation~(\ref{DphatiDphatjfLPT}) is
\begin{equation}
\label{vPapprox}
  \sum_\bk v^\prime_i(\bk) v^\prime_j(-\bk) P(k)
  \approx
  f^\prime_{ij}
  P(2 k_b)
\end{equation}
and equation~(\ref{DphatiDphatjfLPT}) reduces to
\begin{eqnarray}
\label{DphatiDphatjfL}
  \left\langle \Delta\phat^\prime_i(k_1) \Delta\phat^\prime_j(k_2) \right\rangle
  &\approx&
  f^\prime_{ij}
  \bigl[
  \left\langle \Delta\phat(k_1) \Delta\phat(k_2) \right\rangle
\nonumber
\\
&&
  \mbox{}
  +
  4 \, R_a P(k_1) P(k_2) P(2 k_b)
  \bigr]
\end{eqnarray}
with the term on the last line
being the large-scale beat-coupling contribution.

Equation~(\ref{DphatiDphatjfL})
provides the fundamental justfication
for the weightings method
when beat-coupling is taken into account.
It states that
the covariance of shell-averaged power spectra of weighted densities
is proportional to the sum of
the true covariance
$\left\langle \Delta\phat(k_1) \Delta\phat(k_2) \right\rangle$
of shell-averaged power,
and a beat-coupling term
$4 \, R_a P(k_1) P(k_2) P(2 k_b)$
proportional to power at (twice) the box wavenumber $k_b$.
The crucial feature of equation~(\ref{DphatiDphatjfL})
is that the constant of proportionality $f^\prime_{ij}$,
equation~(\ref{fpijv}),
depends only on the weightings $w_i(\bk)$,
and is independent either of
the power spectrum $P(k)$
or of the wavenumbers $k_1$ and $k_2$.

In the limit of infinite box size,
the beat-coupling contribution
to the covariance of power spectra of weighted densities
in equation~(\ref{DphatiDphatjfL})
goes to zero,
$P(2 k_b) \rightarrow P (0) = 0$
as $k_b \rightarrow 0$,
and the covariance becomes proportional
to the true covariance
$\left\langle \Delta\phat(k_1) \Delta\phat(k_2) \right\rangle$
of power.
However, in cosmologically realistic simulations,
such as illustrated in Figure~\ref{vartwofig}
and discussed further in \S\ref{discussion},
the beat-coupling contribution,
far from being small,
is liable to dominate at nonlinear scales.

Beyond perturbation theory or the hierarchical model,
the weightings method remains applicable
just so long as the hierarchical amplitude $R_a$
in equation~(\ref{DphatiDphatjfL})
is independent of the weightings $ij$.
In general,
$R_a$ could be any arbitrary function of $k_1$, $k_2$,
and the box wavenumber $k_b$.

\subsection{Not quite minimum variance weightings}
\label{notquiteweightings}

Section~\ref{weightings}
derived sets of minimum variance weightings
valid when the covariance, and the covariance of covariance,
of power spectra of weighted densities
took the separable forms given by
equations~(\ref{DphatiDphatjfp}) and (\ref{Dphati2Dphatj2gp}).
When beat-scale coupling is included,
the covariance of power,
equation~(\ref{DphatiDphatjfL}),
still takes the desired separable form
(as long as the hierarchical amplitude $R_a$
is independent of the weightings $ij$),
but the covariance of covariance of power
(eq.~(\ref{Dphati2Dphatj2gpL}) of Appendix~\ref{minvarL})
does not.

In Appendix~\ref{minvarL},
we discuss what happens to the minimum variance derivation
of \S\ref{weightings}
when beat-coupling is included.
We argue that
the minimum variance weightings
of \S\ref{theweightings}
are no longer exactly minimum variance,
but probably remain near minimum variance,
and therefore fine to use in practice.

The factor $2$ on the right hand side of equation~(\ref{DXhat2est})
is no longer correct when beat-coupling is included,
but may remain a reasonable approximation.

\section{Discussion}
\label{discussion}

As shown in \S\ref{beatcoupling},
the covariance of nonlinear power
receives beat-coupling contributions from large scales
whenever power is measured from Fourier modes $\rho(\bk)$
that have a finite spread in wavevector $\bk$,
as opposed to being delta-functions at single discrete wavevectors.
Physically,
the large-scale beat-coupling arises
because a product
$\Delta\rho(\bk) \Delta\rho(-\bk{-}\bvarepsilon)$
of Fourier amplitudes of closely spaced wavevectors
couples by nonlinear gravitational growth
to the beat mode $\Delta\rho(\bvarepsilon)$ between them.

The beat-coupling contribution does not appear
when covariance of power is measured from ensembles of
periodic box simulations,
because in that case power is measured
from products of Fourier amplitudes
$\Delta\rho(\bk) \Delta\rho(-\bk)$
at single discrete wavevectors.
Here the ``beat'' mode is the mean mode,
$\bk {-} \bk = \zero$,
whose fluctuation is by definition always zero,
$\Delta\rho(\zero) = 0$.

There is
on the other hand
a beat-coupling contribution
when covariance of power is measured by
the weightings method,
because the Fourier modes of weighted density
are spread over more than one wavevector.

For weightings constructed from combinations of fundamental modes,
as recommended in
\S\ref{weightings},
the covariance of power spectra of weighted densities
receives beat-coupling contributions from power near the box fundamental $k_b$.
The beat-coupling and normal contributions
to the variance
$\left\langle \Delta\phat^\prime_i(k)^2 \right\rangle$
of nonlinear power
are in roughly the ratio
$P(2 k_b) / P(k)$
of power at the box scale
to power at the nonlinear scale,
according to equation~(\ref{DphatiDphatjfL}).

In cosmologically realistic simulations,
box sizes are typically around the range
$10^2$--$10^3 h^{-1} \Mpc$.
This is just the scale at which the
power spectrum goes through a broad maximum.
For example,
in observationally concordant $\Lambda$CDM models,
power goes through a broad maximum
at $k_{\rm peak} \approx 0.016 \, h \, \Mpc^{-1}$
(e.g.\ \citealt{Tegmark04a}),
corresponding to a box size
$4\upi/k_{\rm peak} \sim 800 \, h^{-1} \Mpc$.
Power at the maximum
is about $25$ times
greater than power at the onset
$k_{\rm trans} \approx 0.3 \, h \, \Mpc^{-1}$
of the translinear regime,
$P(k_{\rm peak}) / P(k_{\rm trans}) \approx 25$,
and the ratio
$P(k_{\rm peak}) / P(k)$
of power increases at more nonlinear wavenumbers $k$.

It follows that in cosmologically realistic simulations
the beat-coupling contribution to the covariance of power
is liable to dominate the normal contribution.
This is consistent with the numerical results
illustrated in Figure~\ref{vartwofig}
and discussed by \citet{RH06},
which show that the variance of power
measured by the weightings method
(which includes beat-coupling contributions)
substantially exceeds, at nonlinear scales,
the variance of power measured by the ensemble method
(which does not include beat-coupling contributions).

\subsection{Relevance to real galaxy surveys}
\label{observation}

In real galaxy surveys,
measured Fourier modes inevitably have finite width $|\Delta\bk| \sim 1/R$,
where $R$ is a characteristic linear size of the survey.
The characteristic size $R$
varies from $100 \, h^{-1} \Mpc$ to a few $1000 \, h^{-1} \Mpc$
(an upper limit is set by the comoving horizon distance,
which is about $10^4 \, h^{-1}\Mpc$
in the concordant $\Lambda$CDM model).

It follows that the covariance of nonlinear power
measured in real galaxy surveys
is liable to be dominated not by the ``true'' covariance of power
(the covariance of power in a perfect, infinite survey),
but rather by the contribution from
beat-coupling to power at the scale of the survey.

This means that one must take great care
in using numerical simulations
to estimate or to predict the covariance
of nonlinear power expected in a galaxy survey.
The scatter in power over an ensemble of periodic box simulations
will certainly underestimate the covariance of power
by a substantial factor at nonlinear scales,
because of the neglect of beat-coupling contributions.

A common and in principle reliable procedure
is to estimate the covariance of power of a galaxy survey
from mock surveys ``observed''
with the same selection rules as the real survey
from numerical simulations large enough to encompass the entire survey
(e.g.\ \citealt{CDS01};
\citealt{PB03};
\citealt{Tegmark04a};
\citealt{VMC04};
\citealt{Blaizot05};
\citealt{FSO05};
\citealt{Cole05};
\citealt{Eisenstein05};
\citealt{Park05}).

It is important that numerical simulations
be genuinely large enough to contain a mock survey.
One should be wary about estimating covariance of power
from mock surveys
extracted from small periodic boxes replicated many times
(e.g.\ \citealt{Yang04}),
since such boxes are liable to be missing power at precisely those wavenumbers,
the inverse scale size of the mock survey,
where beat-coupling should in reality be strongest.
Beat-coupling arises from a real gravitational
coupling to large scale modes,
and the simulation from which a mock survey is extracted
must be large enough to contain such modes.

Further,
it would be wrong
to take, say,
a volume-limited subsample of a galaxy survey,
and then to estimate the covariance of power
from an ensemble of periodic numerical simulations
whose size is that of the volume-limited subsample.
A volume-limited subsample of observational data
retains beat-coupling contributions
to the covariance of power,
whereas periodic box simulations do not.

%
%
%
%

\section{Summary}
\label{summary}

This paper falls into two parts.
In the first part,
\S\ref{estimate} and \S\ref{weightings},
we proposed a new method, the weightings method,
that yields an estimate of the covariance of the power spectrum
of a statistically homogeneous and isotropic density field
from a single periodic box simulation.
The procedure is to
apply a set of weightings to the density field,
and to measure the covariance of power
from the scatter in power over the ensemble of weightings.
In \S\ref{estimate}
we developed the formal mathematical apparatus
that justifies the weightings method,
and in \S\ref{weightings}
we derived sets of weightings that
achieve minimum variances estimates of covariance of power.
Section~\ref{recommend}
gives a step-by-step recipe for applying the weightings method.
We recommend a specific set of 52 minimum variance weightings
containing only combinations of fundamental modes.

In the second part of this paper,
\S\ref{beatcoupling} and \S\ref{discussion},
we discuss an unexpected glitch in the procedure,
that emerged from the periodic box numerical simulations
described in the companion paper \citep{RH06}.
The numerical simulations showed that,
at nonlinear scales,
the covariance of power measured by the weightings method
substantially exceeded that measured
over an ensemble of independent simulations.

In \S\ref{beatcoupling}
we argue from perturbation theory
that the discrepancy between the weightings and ensemble methods
arises from ``beat-coupling'',
in which products of closely spaced Fourier modes
couple by nonlinear gravitational growth
to the large-scale beat mode between them.
Beat-coupling is present whenever nonlinear power is measured from Fourier modes
that have a finite spread in wavevector,
as opposed to being delta-functions at single discrete wavevectors.
Beat-coupling affects the weightings method,
because Fourier modes of weighted densities have a finite width,
but not the ensemble method,
because the Fourier modes of a periodic box are delta-functions of wavevector.

As discussed in \S\ref{discussion},
beat-coupling inevitably affects real galaxy surveys,
whose Fourier modes necessarily have a finite width
of the order of the inverse scale size of the survey.
Surprisingly,
at nonlinear scales,
beat-coupling is liable to dominate
the covariance of power of a real survey.
One would have thought that the covariance of power at nonlinear
scales would be dominated by structure at small scales,
but this is not true.
Rather, the covariance of nonlinear power
is liable to be dominated by beat-coupling
to power at the largest scales of the survey.


A common and valid procedure
for estimating the covariance of power from a real survey
is the mock survey method,
in which artificial surveys are ``observed''
from large numerical simulations,
with the same selection rules as the real survey.
It is important that
mock surveys be extracted from genuinely large simulations,
not from many small periodic simulations stacked together,
since stacked simulations miss the large-scale power
essential to beat-coupling.

Finally, it should be remarked that,
although this paper has considered only the covariance
of the power spectrum,
it is likely that,
in real galaxy surveys and cosmologically realistic simulations,
beat-coupling contributions dominate
the nonlinear variance and covariance
of most other statistical measures,
including higher order $n$-point spectra
such as the bispectrum and trispectrum,
and $n$-point correlation functions in real space,
including the $2$-point correlation function.

\section*{Acknowledgements}

We thank Nick Gnedin, Matias Zaldarriaga and Max Tegmark for helpful
conversations, and Anatoly Klypin and Andrey Kravtsov for making the MPI
implementation of ART available to us and for help with its application.
{\sc grafic} is part of the {\sc cosmics} 
package, which was developed by Edmund Bertschinger under NSF grant 
AST-9318185.  The simulations used in this work were performed at the 
San Diego Supercomputer Center using resources provided by the National 
Partnership for Advanced Computational Infrastructure under NSF 
cooperative agreement ACI-9619020.
This work was supported by NSF grant AST-0205981 and by 
NASA ATP award NAG5-10763.

\appendix

\section{Minimum variance weightings and beat-coupling}
\label{minvarL}

This Appendix describes
how the beat-coupling contributions to covariance of power
discussed in \S\ref{beatcoupling}
modify the minimum variance arguments in \S\ref{weightings}.
The conclusion is that the minimum variance weightings
given in \S\ref{theweightings}
are no longer exact minimum variance,
but are probably near minimum variance,
and therefore fine to use in practice.

With no approximations at all,
the covariance of covariance of shell-averaged power spectra
of weighted densities
(with the deviations $\Delta\phat^\prime_i(k)$ in power
being taken relative to the measured rather than the expected mean power,
eqs.~(\ref{Dphati1}) or (\ref{Dphati2})),
takes the generic form
\begin{eqnarray}
\label{Dphati2Dphatj2}
\lefteqn{
  \Bigl\langle
    \bigl[ \Delta\phat^\prime_i(k_1) \Delta\phat^\prime_i(k_2) - \left\langle \Delta\phat^\prime_i(k_1) \Delta\phat^\prime_i(k_2) \right\rangle \bigr]
}
&&
\nonumber
\\
&&
    \times
    \bigl[ \Delta\phat^\prime_j(k_3) \Delta\phat^\prime_j(k_4) - \left\langle \Delta\phat^\prime_j(k_3) \Delta\phat^\prime_j(k_4) \right\rangle \bigr]
  \Bigr\rangle
\nonumber
\\
&=&
  \sum_{\bvarepsilon_1 + \bvarepsilon_2 + \bvarepsilon_3 + \bvarepsilon_4 = \zero}
  v^\prime_i(\bvarepsilon_1) v^\prime_i(\bvarepsilon_2) v^\prime_j(\bvarepsilon_3) v^\prime_j(\bvarepsilon_4)
\\
\nonumber
\lefteqn{
  \ \ 
  \sum_{\bk_1 \in V_{k_1} , ... , \  \bk_4 \in V_{k_4}}
  E(\bk_1 {-} \bk_1^\prime, -\bk_1 {-} \bk_1^{\prime\prime}, ... , \bk_4 {-} \bk_4^\prime, -\bk_4 {-} \bk_4^{\prime\prime})
}
&&
\end{eqnarray}
where
$v^\prime_i(\bk)$ is defined by equations~(\ref{vi}) and (\ref{vpi1}) or (\ref{vpi2}),
the wavevectors $\bvarepsilon_n$ are defined by
\begin{equation}
  \bvarepsilon_n \equiv - \bk^\prime_n {-} \bk^{\prime\prime}_n
  \quad
  \mbox{for $n = 1, ... , 4$}
\end{equation}
and $E$
is an 8-point object,
a sum of products of $n$-point functions
adding up to 8 points,
as enumerated in equation~(\ref{eightpt}).

\eightpointfig

Figure~\ref{eightpointfig}
illustrates the configuration of the $8$-point function
that contributes to the $8$-point object $E$
in equation~(\ref{Dphati2Dphatj2}).
The short legs
$\bvarepsilon_1$, ..., $\bvarepsilon_4$
of the configuration
constitute a tetrahedron,
whose 6 sides generate beat-couplings to large scale.
None of the legs $\bvarepsilon_n$ is zero,
because a zero leg would make zero contribution 
to equation~(\ref{Dphati2Dphatj2}),
since
$v^\prime_i(\zero) = 0$,
equation~(\ref{vpi1}) or, a posteriori, equation~(\ref{vpi2});
but it is possible for the sum of a pair of the short legs to be zero.
It is these configurations,
where the sum of a pair of short legs is zero,
that prevent
equation~(\ref{Dphati2Dphatj2})
from being separated,
as in equation~(\ref{Dphati2Dphatj2gp}),
into a product of
a factor $g^\prime_{ij}$ that depends only on the weightings
and a factor that is independent of the weightings.
Note that it is only the 8-point function itself,
not lower-order functions,
that prevent separability:
lower-order functions depend
on at most three of the four $\bvarepsilon_n$,
and a triangle with three non-zero sides has
(of course) no zero sides.

In either perturbation theory or the hierarchical model,
and in the various valid approximations made in this paper
(firstly, that $n$-point spectra are slowly varying functions
of their arguments
{\em except\/} that
small arguments $\bvarepsilon$ are {\em not\/} replaced by zero,
and secondly, that shells are broad),
the covariance of covariance of shell-averaged power spectra
of weighted densities, equation~(\ref{Dphati2Dphatj2}),
reduces to
\begin{eqnarray}
\label{Dphati2Dphatj2gpL}
\lefteqn{
  \Bigl\langle
    \bigl[ \Delta\phat^\prime_i(k_1) \Delta\phat^\prime_i(k_2) - \left\langle \Delta\phat^\prime_i(k_1) \Delta\phat^\prime_i(k_2) \right\rangle \bigr]
}
&&
\nonumber
\\
&&
    \times
    \bigl[ \Delta\phat^\prime_j(k_3) \Delta\phat^\prime_j(k_4) - \left\langle \Delta\phat^\prime_j(k_3) \Delta\phat^\prime_j(k_4) \right\rangle \bigr]
  \Bigr\rangle
\nonumber
\\
&\approx&
  \lambda g^\prime_{ij}
  - \mu f^\prime_{ii} f^\prime_{jj}
  - \nu f^{\prime \, 2}_{ij}
\end{eqnarray}
where $f^\prime_{ij}$ and $g^\prime_{ij}$
are given by equations~(\ref{fpijv}) and (\ref{gpiju}).
The quantities
$\lambda$, $\mu$, $\nu$
in equation~(\ref{Dphati2Dphatj2gpL})
are each functions of
$k_1, k_2, k_3, k_4$,
and the box wavenumber $k_b$,
but, importantly,
are independent of the weightings $ij$.
The $f^\prime_{ii} f^\prime_{jj}$
(respectively
$f^{\prime \, 2}_{ij}$)
term
in equation~(\ref{Dphati2Dphatj2gpL})
arises from terms
where
$\bvarepsilon_1 {+} \bvarepsilon_2 = \zero$
(respectively
$\bvarepsilon_1 {+} \bvarepsilon_3 = \zero$
or
$\bvarepsilon_1 {+} \bvarepsilon_4 = \zero$)
in equation~(\ref{Dphati2Dphatj2}).
All three terms
on the right hand side of equation~(\ref{Dphati2Dphatj2gpL})
contain large-scale beat-coupling contributions,
proportional to one, two, or three factors of large-scale power.
The second ($\mu$) and third $(\nu)$ terms
in equation~(\ref{Dphati2Dphatj2})
are written with negative signs
because their effect is such as to cancel
some of the beat-coupling terms appearing in the first ($\lambda$) term
(that is,
some of the beat-coupling terms proportional to $P(2 k_b)$ in $\lambda$
should really be proportional to $P(0) = 0$;
the $\mu$ and $\nu$ terms remove these terms).
It is to be expected
that
$\lambda$, $\lambda - \mu$, and $\lambda - \nu$
are all positive.
At linear scales,
where fluctuations are Gaussian,
the beat-couplings generated by nonlinear evolution are small,
so that
$\lambda \gg \mu, \nu$.
At nonlinear scales, however,
$\mu$ and $\nu$ could be an appreciable fraction of $\lambda$.

The derivation of minimum variance weightings in \S\ref{weightings}
involved summing over weighings $ij$,
equation~(\ref{DXhat2}).
Consider the corresponding double sum over $ij$ of
equation~(\ref{Dphati2Dphatj2gpL}).
The sum over $g^\prime_{ij}$
yields the same result as before,
equation~(\ref{DXhat2u}).
Adjoining the sum over $f^\prime_{ii} f^\prime_{jj}$
from equation~(\ref{Dphati2Dphatj2gpL})
modifies equation~(\ref{DXhat2u}) to
\begin{eqnarray}
\label{DXhat2uL}
\lefteqn{
  \frac{1}{(f^\prime N)^2} \sum_{ij}
  \left[
    g^\prime_{ij} - (\mu/\lambda) f^\prime_{ii} f^\prime_{jj}
  \right]
}
\nonumber
\\
&=&
  \frac{1}{u^\prime(\zero)^2}
  \left[
  \left( 1 - \frac{\mu}{\lambda} \right)
  u^\prime(\zero)^2
  +
  \sum_{\bk \neq \zero}
  \left| u^\prime(\bk) \right|^2
  \right]
\end{eqnarray}
because $f^\prime_{ii} = f^\prime = u^\prime(\zero)$ for all $i$.
The minimum variance weightings given in \S\ref{theweightings}
were absolute minimum variance
in the sense that $u^\prime(\bk) = 0$ for $\bk \neq \zero$.
The same minimum variance weightings continue
to achieve absolute minimum variance for equation~(\ref{DXhat2uL}),
reducing its right hand side to the irreducible minimum $1 - \mu/\lambda$.

Thus the minimum variance weightings of \S\ref{theweightings}
remain minimum variance as long as only the first two terms
($\lambda$ and $\mu$)
of equation~(\ref{Dphati2Dphatj2gpL}) are considered.
The third ($\nu$) term
breaks the minimum variance derivation.
However, this third term is likely to be subdominant
compared to the first two.
The quantity $f^\prime_{ij}$
in the third term of equation~(\ref{Dphati2Dphatj2gpL})
is proportional to the covariance of power between weightings $i$ and $j$,
equation~(\ref{DphatiDphatjfL}),
and the Schwarz inequality guarantees that
\begin{equation}
  f^{\prime \, 2}_{ij} \leq f^\prime_{ii} f^\prime_{jj}
\end{equation}
so that there is a natural tendency
for the third term of equation~(\ref{Dphati2Dphatj2gpL})
to be dominated by the second.
The only way for the third term to be large
is for the power spectra from
different weightings $ij$ to be highly correlated with each other.
Physically, however,
the most accurate estimate of covariance of power
should come from averaging over
many uncorrelated weightings,
in which case
$f^{\prime \, 2}_{ij} \ll f^\prime_{ii} f^\prime_{jj}$
for most weightings $i \neq j$.
Thus, as just stated,
it is to be expected that the third term should
be subdominant compared to the first two.

In summary,
to the extent that either
perturbation theory or the hierarchical model
provide a reliable guide to the behaviour
of high-order correlations,
and to the extent that
the third term of equation~(\ref{Dphati2Dphatj2gpL})
is subdominant, as it should be,
the minimum variance weightings of \S\ref{theweightings}
should remain near minimum variance,
good enough for practical application.




\end{document}